\documentclass[11pt, a4paper]{article}

\usepackage[english]{babel}
\usepackage{tikz}
\usepackage[compat=1.1.0]{tikz-feynman}
\usepackage{jheppub}
\usepackage{a4wide}
\usepackage[colorinlistoftodos,textwidth=1.8in]{todonotes}
\usepackage{float}
\usepackage{bbold}
\usepackage{tabularx}
\usepackage{graphicx}
\usepackage{caption}
\usepackage{subcaption}
\usepackage[toc,page]{appendix}
\usepackage[dvipsnames,table]{xcolor}
\usepackage[output-decimal-marker={,},uncertainty-separator = {\pm}, separate-uncertainty = true]{siunitx}
\usepackage{rotating}
\usepackage{amsmath}
\usepackage{verbatim}
\usepackage{tabu}
\usepackage{rotating}
\usepackage{physics}
\usepackage{slashed}
\usepackage[bottom]{footmisc}
\usepackage{amsthm}
\usepackage{csquotes}
\usepackage{multirow}
\usepackage{url}
\usepackage{cleveref}
\usepackage{braket}
\usepackage{gensymb}
\usepackage{amssymb}
\usepackage{placeins}
\usepackage{makecell}
\usepackage{enumitem}
\usepackage[normalem]{ulem}
\newcommand{\ovbb}{$0\nu\beta\beta$\text{ }}

\setlist{itemsep=2em}

\abstract{We study the impact of light GeV-scale heavy neutral leptons (HNLs) on Big Bang nucleosynthesis (BBN) in the neutrino-extended Standard Model Effective Field Theory ($\nu$SMEFT). 
We show that, based on very general considerations, BBN constraints complement laboratory searches at colliders, beam dumps, and neutrinoless double beta decay, by providing an upper bound on the cut-off scale of the effective field theory for HNL masses above $\sim100\,$MeV. 
We identify target regions for future laboratory probes of the $\nu$SMEFT parameter space that is bounded from above and below.}

\begin{document}

\title{Big Bang Nucleosynthesis and the Neutrino-Extended Standard Model Effective Field Theory}

\author[a,b]{Pieter Braat,}\emailAdd{pbraat@nikhef.nl}
\author[a,b]{Jordy de Vries,}\emailAdd{j.devries4@uva.nl}
\author[a,b]{Jelle Groot,}\emailAdd{j.groot4@uva.nl}
\author[d]{Julian Y. Günther,}\emailAdd{jyahguen@uni-bonn.de}
\author[c,a,b]{Juraj Klari\'c}\emailAdd{juraj.klaric@phy.hr}
\affiliation[a]{Institute for Theoretical Physics Amsterdam and Delta Institute for Theoretical
	Physics, University of Amsterdam, Science Park 904, 1098 XH Amsterdam, The
	Netherlands}
\affiliation[b]{Nikhef, Theory Group, Science Park 105, 1098 XG, Amsterdam, The Netherlands}
\affiliation[c]{Department of Physics, Faculty of Science, University of Zagreb,
10000 Zagreb, Croatia}
\affiliation[d]{Bethe Center for Theoretical Physics \& Physikalisches Institut der Universit\"{a}t Bonn, Nu{\ss}allee 12, 53115 Bonn, Germany}
\date{\today}

\maketitle

\thispagestyle{empty}

\setcounter{page}{1}


\section{Introduction}
Neutrino-flavor oscillation experiments~\cite{Super-Kamiokande:1998kpq, SNO:2002tuh, DoubleChooz:2011ymz, DayaBay:2012fng, RENO:2012mkc, T2K:2013ppw} have shown that neutrinos are massive particles. The Standard Model (SM) in its original form cannot accommodate these neutrino masses via the usual Higgs mechanism. 
The SM can be extended by introducing a right-handed (RH) gauge-singlet neutrino that couples to the Higgs and the left-handed (LH) neutrino fields through a Yukawa interaction. 
After electroweak symmetry breaking (EWSB), this coupling generates a Dirac mass term that can account for the non-zero neutrino masses in a renormalizable way. 
Adding a RH neutrino field also allows for a lepton-number-violating (LNV) Majorana mass term that comes with a mass scale that is, in principle, unrelated to the EWSB scale.

The seesaw mechanism~\cite{Minkowski:1977sc, Yanagida:1979as, Gell-Mann:1979vob, Mohapatra:1979ia, Schechter:1980gr, Wyler:1982dd, Mohapatra:1986bd, Bernabeu:1987gr, Akhmedov:1995ip, Akhmedov:1995vm, Malinsky:2005bi} suggests that the neutrino mass eigenstates are Majorana in the presence of both Dirac and Majorana mass terms. 
If the associated mass scale is sufficiently large ($>\,$eV), the smallness of the SM active neutrino masses can be naturally explained by the existence of heavy neutral leptons (HNLs)~\cite{Shrock:1980vy, Shrock:1980ct, Shrock:1981wq, Capozzi:2021fjo, Esteban:2020cvm, deSalas:2020pgw}. 
These particles, often referred to as sterile neutrinos, are much heavier than the active neutrinos. 
The Majorana nature of active neutrinos predicted by the seesaw mechanism allows for processes that violate lepton number by an even number of units. 
The masses of HNLs are not predicted and may span from the eV scale to $10^{15}\,$GeV, motivating scrutiny across all viable mass scales. 
In this work, we focus on HNLs with masses in the MeV-GeV range.  

Over the past decade, there has been sustained interest in such relatively light GeV-scale HNLs (see, e.g., Refs.~\cite{Atre:2009rg, Helo:2013esa, Bondarenko:2018ptm, Drewes:2019fou, Cottin:2022nwp, Liu:2023gpt,Wang:2025esc}).
This interest is driven both by developments in low-scale leptogenesis~\cite{Fukugita:1986hr,Akhmedov:1998qx,Asaka:2005pn,Pilaftsis:2005rv}, and by current and future long-lived particle (LLP) searches at colliders, beam-dump, and neutrino experiments. 
In such experimental searches, MeV-GeV-scale HNLs could be produced via rare decays of abundantly produced mesons. 

In the minimal scenario where HNLs only interact with the SM through active-sterile mixing, stringent bounds have been set on the corresponding mixing angles, hinting that HNLs would be long-lived if they exist~\cite{Deppisch:2015qwa, Bryman:2019bjg, Abdullahi:2022jlv}. 
The decays of sufficiently long-lived HNLs can produce displaced vertices (DVs) that are reconstructible in DV searches. 
These include LHC experiments such as \texttt{ANUBIS}~\cite{Bauer:2019vqk}, \texttt{CODEX-b}~\cite{Gligorov:2017nwh}, \texttt{FACET}~\cite{Cerci:2021nlb}, \texttt{FASER(2)}~\cite{Feng:2017uoz, FASER:2018eoc}, \texttt{MATHUSLA}~\cite{Chou:2016lxi, Curtin:2018mvb, MATHUSLA:2020uve}, \texttt{MoEDAL-MAPP1,2}~\cite{Pinfold:2019nqj, Pinfold:2019zwp}, the beam-dump experiment~\texttt{SHiP}~\cite{SHiP:2015vad, Alekhin:2015byh, SHiP:2018xqw, SHiP:2021nfo}, as well as the \texttt{DUNE} near detector (\texttt{DUNE-ND})~\cite{DUNE:2020lwj, DUNE:2020jqi, DUNE:2020ypp,  
DUNE:2021mtg, DUNE:2021cuw} at Fermilab's Long-Baseline Neutrino Facility (\texttt{LBNF}). 

In numerous BSM scenarios beyond the minimal case, HNLs are feebly interacting particles at low energies. 
At higher energy scales, such HNLs acquire interactions through the exchange of heavy BSM mediators. 
Representative examples that introduce new heavy fields above the electroweak scale include left-right symmetric theories \cite{Mohapatra:1979ia, Pati:1974yy, Mohapatra:1974gc, Senjanovic:1975rk, Mohapatra:1980yp, Senjanovic:1978ev}, $Z^\prime$ scenarios~\cite{Deppisch:2019kvs, Chiang:2019ajm}, and leptoquark models~\cite{Cottin:2021tfo, Bhaskar:2023xkm}. 
In such scenarios, regardless of the underlying model details, HNL dynamics at low energies can be systematically described by local effective operators within the neutrino-extended Standard Model Effective Field Theory ($\nu$SMEFT)~\cite{delAguila:2008ir, Aparici:2009fh, Bhattacharya:2015vja, Liao:2016qyd}.

The $\nu$SMEFT framework is general and therefore quite powerful. 
However, this generality makes it difficult to identify well-defined target regions for experiments, even for relatively light HNLs. 
In the minimal scenario, by contrast, the requirement of generating neutrino masses sets a lower bound on the HNL coupling strength, and the additional requirement of successful leptogenesis further restricts the allowed parameter space, rendering the model highly testable (see, e.g., Ref.~\cite{Klaric:2020phc, Drewes:2021nqr,Hernandez:2022ivz,deVries:2024rfh,Fuyuto:2025feh,Chun:2017spz}). 
Conversely, existing $\nu$SMEFT analyses, such as those in Refs.~\cite{Cottin:2021lzz, Beltran:2021hpq, Fernandez-Martinez:2023phj, Beltran:2025ilg}, typically set only lower bounds on the NP scale $\Lambda$ for a given HNL mass. 
While these bounds can correspond to impressive sensitivities, reaching up to hundreds of TeV in DV searches or neutrinoless double beta decay ($0\nu\beta\beta$)~\cite{Cirigliano:2017djv, Dekens:2020ttz}, they can always be evaded by increasing $\Lambda$.

Cosmological probes can provide important complementary information. 
The thermal interaction rates of HNLs induced by the $\nu$SMEFT operators grow with temperature, leading to their thermalization in the early universe, assuming sufficiently high reheating temperatures. 
In this case, the HNLs affect the expansion rate of the universe and, depending on their lifetimes, can strongly influence big bang nucleosynthesis (BBN)~\cite{Steigman:2012ve, Cyburt:2015mya, Kawasaki:2017bqm, Pitrou:2018cgg, Pospelov:2010hj, Dolgov:2000jw, Ruchayskiy:2012si, Boyarsky:2020dzc, Sabti:2020yrt} or the cosmic microwave background (CMB)~\cite{Hernandez:2013lza, Hernandez:2014fha, Vincent:2014rja}.
With rather mild assumptions -- most notably the reheating temperature requirement -- an \textit{upper bound} on $\Lambda$ can be derived from cosmology that complements the lower bound obtained from laboratory experiments. 
As a result, well-defined target regions can be identified for future experiments, specified by the HNL mass and the $\nu$SMEFT scenario. 
The main goal of this paper is to derive these target regions for HNLs with masses above 100 MeV, which are the primary focus of upcoming experiments such as \texttt{DUNE} and \texttt{SHiP}. 

This paper is organized as follows. In Sec.~\ref{sec:EFT}, we introduce the $\nu$SMEFT framework. 
In Sec.~\ref{sec:3_EarlyUniverseDecay}, we discuss HNLs in the early universe and how $\nu$SMEFT operators can impact BBN. 
In Sec.~\ref{sec:ovbb} and \ref{sec:DV} we discuss laboratory constraints from, respectively, $0\nu\beta\beta$ and DV searches. 
In Sec.~\ref {sec:Results}, we discuss several $\nu$SMEFT scenarios and show the complementarity of laboratory searches and BBN considerations. 
We conclude in Sec.~\ref{sec:conclusions}.

\section{Standard model effective field theory extended by HNLs}
\label{sec:EFT}
We focus on GeV-scale HNLs that are kinematically accessible at collider experiments. 
To accommodate such states, we extend the SM field content by introducing RH neutrino fields $\nu_R$. 
These fields are singlets under the SM gauge group and are therefore often referred to as sterile neutrinos. 
At the renormalizable level, they couple exclusively to the SM through Yukawa couplings with the Higgs field and lepton doublets.  
These interactions define the neutrino portal. 

\subsection{The effective neutrino Lagrangian}

The renormalizable interactions in this framework are 
\begin{align}
\label{eq:LnuRdim4}
    \mathcal{L}=-\left[\frac{1}{2}\bar{\nu}^c_{R,i}\bar{M}_{R,ij}\nu_{R,j}+\epsilon_{\alpha\beta}L_{\alpha,k}^T{\tilde H}_\beta Y_{\nu,kj}\nu_{R,j}+\text{h.c.}\right]+\mathcal{L}_{\rm SM}\ ,
\end{align}
where the Majorana mass term of the RH neutrino contains the $n\times1$ column vector $\nu_R$ with generation indices $i,j\in {1,2,\dots,n}$, and the complex $n\times n$ symmetric Majorana mass matrix $M_R$. 
The neutrino portal involves the anti-symmetric $SU(2)_L$ tensor $\epsilon$ with $\alpha,\beta\in 1,2$ and $\epsilon_{12}=+1$, the SM lepton doublet $L$ with generation index $k\in{1,2,3}$, the complex Higgs doublet ${\tilde H}\equiv i\tau_2 H$, and the $3\times n$ Yukawa-coupling matrix $Y_\nu$. 
In the unitary gauge, the Higgs doublet is given by $H=v/\sqrt{2}(0, 1+h/v)^T$, where $h$ denotes the Higgs boson and $v=246\,$GeV is the Higgs vacuum expectation value (vev). 
$\mathcal{L}_{\rm SM}$ is the SM Lagrangian. 
For any field $\psi$, its charge conjugate is defined through $\psi^c\equiv C\bar{\psi}^T$, where $C=-i\gamma^2\gamma^0$ is a charge conjugation matrix that satisfies $C=-C^{-1}=-C^T=-C^\dagger$. 
The chiral spinors $\psi_{L,R}\equiv P_{L,R}\psi$ are defined using the chiral projection operators $P_{L,R}=(1\mp\gamma^5)/2$, where $\psi^c_{L,R}\equiv (\psi_{L,R})^c = C \overline{(P_{L,R}\psi)}^T=P_{R,L}\psi^c$. 

If charged under non-SM gauge symmetries, HNLs could interact with particles that are decoupled at low energies, maintaining the validity of the label `sterile'. 
A popular example is the minimal left-right-symmetric model~\cite{Mohapatra:1979ia, Pati:1974yy, Mohapatra:1974gc, Senjanovic:1975rk, Mohapatra:1980yp, Senjanovic:1978ev}, in which RH neutrinos are charged under an $SU(2)_R$ gauge symmetry and couple to RH gauge bosons, as well as additional scalar fields. 
Such new mediators must be significantly heavier than the electroweak (EW) scale $v$ to avoid experimental constraints, whereas the HNLs could still be light. This separation of scales suggests the use of EFT techniques. 

In a SM EFT framework extended by RH neutrino fields ($\nu$SMEFT), the Lagrangian in Eq.~\eqref{eq:LnuRdim4} is the renormalizable part ($\leq$ dim-$4$) of a tower of operators described by 
\begin{align}
    \mathcal{L}_{\nu \text{SMEFT}}=\mathcal{L}+\sum_{d>4}\sum_i C_i^{(d)}\mathcal{O}_i^{(d)}\ ,
\end{align}
where the $d$-dimensional operators $\mathcal{O}_i^{(d)}$ are scaled by Wilson coefficients $C_i^{(d)}$ that are suppressed by powers of $\Lambda^{4-d}$, where $\Lambda(\gg v)$ is the matching scale of the UV theory. 

We focus on operators of dimension six at most and consider only one HNL ($n=1$) for simplicity. 
In this limit, the only two relevant dimension-5 operators are
\begin{align}
    \mathcal{L}^{(5)}_{\nu_L}= \epsilon_{ij}\epsilon_{kl}L^T_iC^{(5)}_{\nu_L}CL_kH_jH_l\ , \quad  \mathcal{L}_{\nu_R}^{(5)}=-\bar{\nu}_R^cM_R^{(5)}\nu_RH^\dagger H\ ,
\end{align}
which contribute to the Majorana mass terms of the active and sterile neutrinos after EWSB. 
The contributions of these operators can be absorbed into the neutrino masses, except for those involving interactions with Higgs bosons, which do not enter processes of interest. 

For dimension-6 operators, we focus on interactions that include at most one HNL field. All relevant operators in this limit are tabulated in Tab.~\ref{tab:dim6operators}. 
After EWSB, the operator $\mathcal{O}^{(6)}_{L\nu H}$ generates corrections to the Yukawa couplings $Y_{\nu,ij}$ in Eq.~\eqref{eq:LnuRdim4} and induces operators with explicit Higgs bosons, which are phenomenologically irrelevant. 
The interesting physics of neutrino magnetic moments induced by the dipole operators $\mathcal{O}^{(6)}_{\nu W}$ and $\mathcal{O}^{(6)}_{\nu B}$ has been discussed recently in Refs.~\cite{Brdar:2020quo, Fuyuto:2024oii}. 
We focus here on $\mathcal{O}^{(6)}_{\nu W}$, which induces $0\nu\beta\beta$ at tree level at lower energies \cite{Dekens:2020ttz}. 
In addition, we consider the five four-fermion operators and the class-2 operator $\mathcal{O}^{(6)}_{H\nu \ell}$.

\begin{table}[t!]
    \centering
    \renewcommand{\arraystretch}{1.5}
    \begin{tabular}{|cc||cc|}
        \hline
        \multicolumn{2}{|c||}{\textbf{Class 1~:$\quad \psi^2 H^3$}} & \multicolumn{2}{c|}{\textbf{Class 4~:$\quad \psi^4$}} \\
        \hline
        $\mathcal{O}^{(6)}_{L\nu H}:$ & $(\bar{L} \nu_R) \tilde{H} (H^\dagger H)$ & $\mathcal{O}^{(6)}_{du\nu \ell}:$ & $(\bar{d}_R \gamma^\mu u_R)(\bar{\nu}_R \gamma_\mu \ell)$ \\
        \cline{1-2}
        \multicolumn{2}{|c||}{\textbf{Class 2~:$\quad \psi^2 H^2 D$}} & $\mathcal{O}^{(6)}_{QuvL}:$ & $(\bar{Q} u_R)(\bar{\nu}_R L)$ \\
         \cline{1-2}
        $\mathcal{O}^{(6)}_{H\nu \ell}:$ & $(\bar{\nu}_R \gamma^\mu \ell)(\tilde{H}^\dagger i D_\mu H)$ & $\mathcal{O}^{(6)}_{L\nu Qd}:$ & $(\bar{L} \nu_R) \epsilon (\bar{Q} d)$ \\
        \cline{1-2}
        \multicolumn{2}{|c||}{\textbf{Class 3~:$\quad \psi^2 H^3 D$} }& $\mathcal{O}^{(6)}_{LdQ\nu}:$ & $(\bar{L} d_R) \epsilon (\bar{Q} \nu_R)$ \\
         \cline{1-2}
        $\mathcal{O}^{(6)}_{\nu W}:$ & $(\bar{L} \sigma_{\mu \nu} \nu_R) \tau^I \tilde{H} W^{I \mu \nu}$ & $\mathcal{O}^{(6)}_{L\nu L\ell}:$ & $(\bar{L} \nu_R) \epsilon (\bar{L} \ell_R)$ \\
        $\mathcal{O}^{(6)}_{\nu B}:$ & $(\bar{L} \sigma_{\mu \nu} \nu_R) \tilde{H} B^{ \mu \nu}$ & \multicolumn{2}{c|}{}\\
        \hline
    \end{tabular}
    \caption{Classification of all dimension-six operators in the $\nu$SMEFT framework that involve one HNL field \cite{Liao:2016qyd}.}
    \label{tab:dim6operators}
\end{table}

Using one-loop QCD anomalous dimensions, we evolve these operators from the UV scale $\Lambda$ to the electroweak scale $v$. 
The relevant RGEs are obtained as \cite{Dekens:2020ttz}
\begin{align}
\label{eq:RGEs}
    \frac{dC^{(6)}_{Qu\nu L}}{d\ln \mu}=\frac{\alpha_s}{4\pi}\gamma_S C^{(6)}_{Qu\nu L}\ ,\quad \frac{dC^{(6)}_{S}}{d\ln \mu}=\frac{\alpha_s}{4\pi}\gamma_S C^{(6)}_{S}\ ,\quad \frac{dC^{(6)}_{T}}{d\ln \mu}=\frac{\alpha_s}{4\pi}\gamma_T C^{(6)}_{T}\ ,
\end{align}
where we have defined 
\begin{align}
    C_S^{(6)}\equiv -\frac{1}{2}C_{LdQ\nu}^{(6)}+C^{(6)}_{L\nu Q d}\ ,\quad C_T^{(6)}\equiv-\frac{1}{8}C_{LdQ\nu}^{(6)}\ ,
\end{align}
where $\gamma_S=-6C_F$ and $\gamma_T=2C_F$ with the Casimir factor $C_F=(N_c^2-1)/(2N_c)$, where $N_c=3$ is the number of colors. 
The vector-current operators $\mathcal{O}^{(6)}_{H\nu \ell}$ and $\mathcal{O}^{(6)}_{du\nu \ell}$ do not evolve under QCD at the one-loop level.  

The heavy degrees of freedom -- namely the bosons $H, W^\pm$, $Z$, and the top quark $t$ -- can be integrated out at the electroweak scale. 
The remaining Lagrangian can be matched to a RH neutrino-extended low-energy EFT ($\nu$LEFT) that is invariant under $SU(3)_C\otimes U(1)_{\rm EM}$. 
The relevant dimension-6  charged-current contributions to the $\nu$LEFT Lagrangian are 
\begin{align}
\label{eq:nuleftCC}
\notag \mathcal{L}^{(6){\rm CC}}_{\nu{\rm LEFT}} = &-\frac{2G_F}{\sqrt{2}}\sum_{i,j}{\Big\{}2[\bar{\nu}_L^i\gamma_\mu \ell_L^i][\bar{\ell}_L^j\gamma_\mu \nu_L^j]-\sum_k{\Big [}-2V_{ij}[\bar{u}^i_L\gamma^\mu d_L^j][\bar{\ell}^k_L\gamma_\mu \nu_L^k]\\& \notag +{\Big (}c^{{\rm CC}}_{{\rm VLR}1,ijk}[\bar{u}^i_L\gamma^\mu d_L^j]+c^{\rm CC}_{{\rm VRR},ijk}[\bar{u}_R^i\gamma^\mu d_R^j]{\Big )}[\bar{\ell}_R^k\gamma_\mu \nu_R]\\
&+{\Big (}c^{\rm CC}_{{\rm SRR},ijk}[\bar{u}^i_Ld^j_R]\notag+c^{\rm CC}_{{\rm SLR},ijk}[\bar{u}_R^id_L^j]{\Big )}[\bar{\ell}_L^k\nu_R] +c_{{\rm T},ijk}^{\rm CC
}[\bar{u}_L^i\sigma^{\mu\nu}d_R^j][\bar{\ell}_L^k\sigma_{\mu\nu}\nu_R]\\
&+c^{\rm CC}_{{\rm VLR}2,ijk}[\bar{\nu}_L^i\gamma^\mu \ell_L^j][\bar{\ell}_R^k\gamma_\mu\nu_R]+\text{h.c.}{\Big ]}{\Big \}}\ ,
\end{align}
where $f_{L,R}$ for $f\in u,d,\ell,\nu$ are chiral up-type quark, down-type quark, charged lepton and neutrino fields, respectively, and $i,j,k\in 1,2,3$ denote generation indices.
The relevant neutral-current interactions are 
\begin{align}
\label{eq:nuleftNC}
    \notag \mathcal{L}^{(6){\rm NC}}_{\nu {\rm LEFT}} = &-\frac{2G_F}{\sqrt{2}} \sum_{i,j} {\Big \{}[2\eta_{Z}^{ij}(f) c^{\rm NC}_{Z}({f})[\bar{\nu}_L^i\gamma_\mu \nu_L^i] [\bar{{f}}^j\gamma^\mu {f}^j]-\sum_{k}{\Big [}{\Big (}c^{\rm NC}_{{\rm SLR},ijk}[\bar{u}_R^iu_L^j]\\&+c_{{\rm SRR},ijk}^{\rm NC}[\bar{d}_L^id_R^j] {\Big )} [\bar{\nu}_L^k\nu_R]+c^{\rm NC}_{{\rm T},ijk}[\bar{d}_L^i\sigma^{\mu\nu}d_R^j][\bar{\nu}_L^k\sigma_{\mu\nu}\nu_R]+\text{h.c.}{\Big ]}{\Big \}}\ ,
\end{align}
with chiral fermion fields $f\in u_L,u_R,d_L,d_R,\ell_L,\ell_R,\nu_L,\nu_R$ and the weak neutral current coupling $c^{\rm NC}_{Z}({f})=T^3_L-\sin^2 \theta_W Q$
with electric charge $Q$, the electroweak mixing angle $\theta_W$, and the isospin projection $T^3_L=\tau^3/2$ for left-handed doublets. 
For the symmetry factor, we have  $\eta_{Z}^{ij}({\nu_L}) =\frac{1}{2}(2-\delta_{ij})$ and $\eta_{Z}^{ij}({f}) =1$ otherwise. 

For the $\nu$LEFT Wilson coefficients, tree-level matching of the $\nu$SMEFT to the $\nu$LEFT Lagrangian at the electroweak scale yields the charged-current contributions
\begin{align}
\begin{split}
\label{eq:nuLEFTCCs}
   c^{\rm CC}_{{\rm VLR}1,ijk}&=-V_{ij}v^2C_{H\nu \ell,k}^\dagger -  V_{ij}\frac{4 \sqrt{2}v m_l}{g} C_{\nu W \ell,k}\ ,\hspace{2.0em} c^{\rm CC}_{{\rm VLR}2,ijk}=-\delta_{ij}v^2C_{H\nu \ell,k}^\dagger\ , \\  
   c^{\rm CC}_{{\rm VRR},ijk}&=v^2C_{du\nu \ell,jik}^\dagger\ ,\hspace{13em}  c^{\rm CC}_{{\rm SLR},ijk}=V_{lj}v^2C_{Qu\nu L,lik}^\dagger \ ,\\ 
   c^{\rm CC}_{{\rm SRR},ijk}&=-v^2C_{L\nu Qd,kij}+\frac{v^2}{2}C_{LdQ\nu,kji}\ ,\hspace{6.5em} c_{{\rm T},ijk}^{\rm CC}=\frac{v^2}{8}C_{LdQ\nu,kji} \ ,
\end{split}
\end{align}
where we sum over the repeated flavor index $l\in 1,2,3$. 
At low energies, the operators $C_{\nu W l}$ and $C_{H \nu l}$ both contribute to the same $\nu$LEFT operator, but the contribution from the latter is suppressed by $m_l/v \ll 1$. 
For the neutral-current Wilson coefficients, we find
\begin{align}
\begin{split}
\label{eq:nuLEFTNCs}
    c^{\rm NC}_{{\rm SLR},ijk}&=v^2C_{Qu\nu L,jik}^\dagger\ , \quad c^{\rm NC}_{{\rm SRR},ijk}=v^2V_{li}^*C_{L\nu Qd,klj}-\frac{v^2}{2}V_{li}^*C_{LdQ\nu,kjl}\ ,\\ c^{\rm NC}_{{\rm T},ijk}&=-\frac{v^2}{8}V_{li}^*C_{LdQ\nu,kjl}\ .
\end{split}
\end{align}

Employing the running of the one-loop QCD anomalous dimensions, the operators are finally evolved from the EW scale $v$ down to the chiral symmetry-breaking scale $\lambda{\rm \chi} \sim 1$ GeV. 
At this order, only scalar and tensor currents are affected by the running and evolve analogously to $C^{(6)}_{S, T}$ in Eq.~\eqref{eq:RGEs}. 
Numerically, the scalar currents are enhanced by approximately $60\%$, whereas tensor currents are suppressed by about $15\%$ \cite{Liao:2020roy}. 
Although the $\nu$LEFT Lagrangian conserves lepton number at dimension six, there are still contributions to lepton-number-violating processes like \ovbb through Majorana mass insertions in neutrino exchange. 
We discuss this mechanism in more detail in Sec.~\ref{sec:ovbb}.

\section{BBN bounds on HNLs}
\label{sec:3_EarlyUniverseDecay}
Big Bang nucleosynthesis (BBN) is one of the cornerstones of modern cosmology that acts as a probe of the very early moments in the thermal history of the universe, when temperatures were as high as a few MeV.
The abundances of primordial elements produced by this process largely agree with SM predictions and therefore serve as an excellent probe for the existence of any BSM particles that may have been present during this era.

This cosmological constraint is particularly interesting in the context of searches for HNLs,
as it provides a time scale $\tau_\text{BBN} \sim 0.1\,$s~\cite{Sabti:2020yrt,Boyarsky:2020dzc} that can be translated into an effective \emph{upper} bound on the lifetimes of the HNLs. 
One example of the success of such bounds on the properties of HNLs is the minimal seesaw model, where the combination of direct searches and BBN bounds severely limits the allowed masses and mixings of the HNLs~\cite{Bondarenko:2021cpc}. 

In this section, we will discuss how these bounds can be understood in the more general framework of $\nu$SMEFT, as the effective operators can significantly affect both the production and decays of the HNLs, leading to upper bounds on the scale of new physics.

\subsection{Bounds on BBN-stable HNLs}
The presence of HNLs with masses above a few MeV can significantly alter the thermal history of the early universe, even if they do not decay during BBN.
If these HNLs remain stable, they will eventually dominate the total energy density since their energy density redshifts as $\rho_N \propto a^{-3}$, which is slower than radiation $\rho_r\propto a^{-4}$.
In the most extreme example, this leads to an early matter-dominated era, which increases the Hubble rate during BBN,
modifying the freeze-out of weak interactions and the neutron-to-proton ratio, and consequently spoiling the successful predictions of light-element abundances (see e.g.~\cite{Yeh:2024ors} and references therein).
This condition sets an upper bound on the HNL energy density
\begin{align}
\label{eq:simplebound}
    Y_N M_4 \lesssim \frac{3}{4} T_{\rm NS}\ ,
\end{align}
where $T_{\rm NS} \approx 80$ keV is the temperature corresponding to the freeze-out of the thermonuclear reactions at the end of BBN, and $Y_N = n_N/s$ is the HNL yield.

The simple analytic estimate that leads to the bound in Eq.~\eqref{eq:simplebound} is shown in appendix \ref{app:MatterDomination}. 
If we assume that the HNL abundance freezes out while they are still relativistic ($Y_N^f \sim 10^{-3}$), this corresponds to a bound $M_4 \lesssim 30$ MeV.
For masses below this bound, the HNL energy density is too small to directly affect SM BBN.\footnote{However, stable HNLs with $Y_N^f \sim 10^{-3}$ are still incompatible with cosmology for $M_4 \gg \mathcal{O}(\mathrm{keV})$ as they would act as a dark matter component exceeding the observed density~\cite{Planck:2018vyg}.}
We verified that this simple analytic estimate shows rough agreement with numerical codes such as \texttt{AlterBBN}~\cite{Arbey:2018zfh}. 

While this bound appears strong, it can easily be avoided if the HNL abundance is sufficiently low, either because it is low at freeze-out or because the HNLs decay early enough to be largely depleted before the onset of BBN.

\subsection{Bounds on HNL decays}
If HNLs decay sufficiently early, the problem of early matter domination is evaded. 
The most conservative bound is set by requiring that the HNLs decay before the onset of BBN at roughly $\tau_{\rm BBN} \sim 0.1\,$s. 
However, depending on the decay mode of the HNL, this bound can be more stringent \cite{Sabti:2020yrt}. 

We distinguish between three HNL decay modes: decays into mesons (pions), EM final states (electrons/positrons/photons), and neutrinos.
The decay modes will depend on specific operators in Eqs.~\eqref{eq:nuleftCC}, as well as the HNL mass. 
Since the reliability of our analysis diminishes at low HNL masses, as discussed later in this section, we focus on HNL masses above $100\,$MeV.

Several operators in Eq.~\eqref{eq:nuleftCC} induce HNLs decays to charged pions. 
If HNLs indeed decay into charged pions, the neutron-to-proton ($n\leftrightarrow p$) interaction rate is enhanced, which keeps the  $n\leftrightarrow p$ ratio in equilibrium longer compared to standard BBN. 
This increases the number of neutrons available for light-element production, thereby increasing the abundance of helium-4. 
The resulting bound from the observed $\mathcal{Y}_p$ fraction is $\tau_{\rm BBN} \simeq 0.023\,$s \cite{Boyarsky:2020dzc}. 
The exact lifetime bound has a moderate dependence on the HNL yield at freeze-out $Y_N^f = n_N^f/s \approx 10^{-3}$ and the HNL branching fraction into final state mesons ${\rm Br}_{N_4\to h}$ for $h$ $\in \{ \pi^-, K_L^0, K^- \}$. 
This bound is given by \cite{Boyarsky:2020dzc,Bondarenko:2021cpc,Syvolap:2021yan}
\begin{align}
\label{eq:BBNlifetimelim}
    \tau_{N} < \tau_{\mathrm{BBN}} &= \frac{0.023\,\text{s}}{1+0.07\ln \left(\dfrac{P_{\rm conv,{h}}}{0.1}\dfrac{{\rm Br}_{N_4\to h}}{0.1}\dfrac{Y_N^f\zeta}{10^{-3}}\right)}\ ,\\
    \label{eq:simplifiedBBNlifetime}
    &\approx \frac{0.023\,\text{s}}{1+0.07\ln \left(\dfrac{P_{\rm conv,{h}}}{0.1}\dfrac{{\rm Br}_{N_4\to h}}{0.1}\right)}\ ,
\end{align}
where $P_{{\rm conv},h}\sim(10^{-2}-10^{-1})$ for $T=\mathcal{O}(\text{MeV})$ quantifies the probability that a final-state hadron converts a nucleon before decaying.

The dilution factor $\zeta\equiv\left(a_{\rm SM}/a_{\text{SM} + N}\right)^3<1$ accounts for the difference in the scale factor with and without the HNLs present.
For HNLs in the GeV mass range, this ranges from $\zeta \in [0.1, 1]$~\cite{Boyarsky:2020dzc}.
For $\nu$SMEFT operators that contain light quarks and HNLs that freeze out relativistically, the argument of the logarithm in the numerator is at most $\mathcal O(10)$, which mildly decreases the value of $\tau_{\mathrm{BBN}}$ by $\mathcal O(20\%)$.
For HNLs that freeze out non-relativistically, we have to ensure that $Y_N^f \zeta$ has not been reduced too much by Boltzmann suppression at the freeze-out temperature, as this would imply weaker constraints on the lifetime.
Due to the non-trivial relationship between $Y_N^f$ and the Wilson coefficients, we use a simplified BBN bound~\eqref{eq:simplifiedBBNlifetime}, where we assumed $Y_N^f \zeta \gtrsim 10^{-3}$.
We verify this in more detail in an explicit example in Sec.~\ref{sub:example}, and show that this assumption is always valid in the HNL mass range of our interest. 

Alternatively, the $\mathcal{O}_{L\nu Le}$ operator in Eq.~\eqref{eq:nuleftCC} produces additional (left-handed) neutrinos through decays. 
These induce spectral distortions in the (active) neutrino spectrum, which enhance the $p\to n$ rate relative to the $n \to p$ rate. 
Again, if HNLs decay too late, this would lead to an overproduction of helium-4. 
The resulting lifetime bound is slightly weaker than the above-stated meson bound, $\tau_{\rm BBN} \sim 0.03-0.05\,$s \cite{Sabti:2020yrt}. 
This bound is present for all operators in Tab.~\ref{tab:dim6operators}.

\subsection{The thermal history of HNLs}
HNL production occurs over a wide range of temperatures. 
Before the electroweak crossover, their dynamics are described within the full $\nu$SMEFT framework, whereas after the crossover, they can be effectively treated in the $\nu$LEFT framework. 
The production of HNLs in the early universe can generally be described by the Boltzmann equation
\begin{align}
\label{eq:sterile_production}
    z \frac{d Y_N}{d z} =
    - \frac{\Gamma_N}{\mathcal{H}} (Y_N - Y_N^\mathrm{eq})\ ,
\end{align}
where $z=M_4/T$ is the conveniently chosen time variable, $Y_N \equiv n_N/s$  and $Y_N^\mathrm{eq}$ are the HNL yield and equilibrium yield, respectively. 
The effective Hubble rate is given by $\mathcal{H} \approx T^2/M_\mathrm{Pl}^\star$, with additional details included in \cref{sec:BE_sterile}.
The HNL equilibration rate $\Gamma_N$ includes all processes that bring the HNLs into equilibrium, be it decays, inverse decays, or scatterings.
This rate crucially depends on the operators included in the effective Lagrangian.
For dimension-6 operators that involve a single HNL, based on naive dimensional analysis, we can generically expect the scaling
\begin{align}
\label{eq:scaling}
    \Gamma_N \propto
    	\begin{cases}
		\frac{T^5}{\Lambda^4} \text{ in the relativistic limit, for } T>M_4\ ,\\
		\frac{M_4^5}{\Lambda^4} \text{ in the non-relativistic case, for } M_4>T\ .\\
	\end{cases}
\end{align}
In the relativistic limit, this effectively captures the HNL production rate, whereas in the non-relativistic limit, it primarily corresponds to decays.
When $\Gamma_N/\mathcal{H} \gtrsim 1$, the HNL yield closely follows the equilibrium distribution $Y_N \approx Y_N^\mathrm{eq}$.
Because the scattering rate in Eq.~\eqref{eq:scaling} increases faster with temperature than the Hubble rate, HNLs are in equilibrium at high temperatures (assuming a sufficiently high reheating temperature), but gradually freeze out as the universe cools down.

Freeze-out happens at a temperature $T_f$, which corresponds to $\Gamma_N = \mathcal{H}$.
The HNL yield is then approximately frozen at $Y_N^f \simeq Y_N^\mathrm{eq}(T_f)$, until their decays become efficient at a time $t \sim \tau_N =\Gamma_N^{-1}$.

We can estimate the freeze-out abundance based on the HNL production rate in the relativistic limit.
The HNL production typically occurs before quark confinement, allowing us to treat quarks as free particles.
Specifically, for vector-like couplings and a typical HNL momentum $k = \pi T$, we can estimate the production rate as \cite{Ghiglieri:2016xye}
\begin{align}
\label{eq:GammaProd}
 \Gamma^{\rm prod}_{N}= \sum_i 0.3 A_i(T) |c_i|^2 G_F^2 T^5\ ,
\end{align}
where $c_i$ is the corresponding $\nu$LEFT coefficient, while the coefficient $A_i$ captures the remaining factors that enter the rate. 
In the minimal coupling scenario, we can extract the common factor $|c_0|^2 = 4 \sum_l |U_{l4}|^2$, with the remaining coefficient \cite{Ghiglieri:2016xye}
\begin{align}
\label{eq:AMikko}
    A_0 &\approx
    \underset{\text{leptons}}{\underbrace{\frac{15}{2}-2s^2+12s^4}}+N_c\left[\underset{\text{neutral quark currents}}{\underbrace{\frac{5}{2}-\frac{14s^2}{3}+\frac{44s^4}{9}}}+\underset{\text{charged quark currents}}{\underbrace{2\left(|V_{\rm ud}|^2+|V_{\rm us}|^2+|V_{\rm cd}|^2+|V_{\rm cs}|^2\right)}}\right] \simeq 24.7\ ,
\end{align}
where $s=\sin \theta_W$, with the Weinberg angle $\theta_W$ and contributions from third-generation quarks are neglected. 
For general $\nu$SMEFT scenarios, $|c_i|^2 G_F^2 \rightarrow \Lambda^{-4}$ and $A_i$ needs to be determined, as illustrated by an explicit example below.
Since we are interested in HNL production, we typically consider temperatures above the QCD scale ($T > \Lambda_{\rm QCD}$), where quarks are free, and hadronization effects are absent.

The scaling of the production rate gives a simple estimate for the freeze-out temperature
\begin{align}
    T_f = \left( 1.2 A_i |c_i|^2 G_F^2 M_{\rm Pl}^\star \right)^{-1/3}\ .
\end{align}
For a general $\nu$SMEFT scenario, we write
\begin{align}
    \label{eq:TfLambda}
    T_f =\Lambda  \left(\frac{\Lambda}{1.2 A_i M_{\rm Pl}^\star}\right)^{1/3}\ ,
\end{align}
where $T_f \ll \Lambda$ since $\Lambda \ll M_{Pl}^\star$ in experimentally interesting scenarios.
Crucially, this means we can compute the freeze-out temperature within $\nu$SMEFT without specifying the UV completion.\footnote{
    However, for a reheating temperature $T_{\rm reh} < T_f$, the
    HNLs do not reach thermal equilibrium and are highly sensitive to the thermal history of the universe.
}
We can now estimate the freeze-out abundance as
\begin{align}
    \label{eq:YfTf}
    Y_N^f &\simeq Y_N^\mathrm{eq} (T_f)\\
    \label{eq:Boltzmann}
        &\approx
    \begin{cases}
        1.95 \cdot 10^{-3} & \text{for } M_4 \ll T_f\ ,\\
        1.35 \cdot 10^{-3} e^{-M_4/T_f} \left( \frac{M_4}{T_f} \right)^{3/2} & \text{for } M_4 \gg T_f\ ,\\
    \end{cases}
\end{align}
which is either set by relativistic abundances in the relativistic freeze-out case, or by Boltzmann suppression in the non-relativistic case.
We have assumed $g_\star = 106.75$, which is only true above the EW crossover $T_f \gtrsim 100$ GeV. 
This leads to a slight underestimate of the yield for HNLs with $T_f = \mathcal{O}({\rm MeV})$.
As we will see in the following section, for HNLs with $M_4 \gtrsim m_\pi$ that satisfy the lifetime bound, it is typically safe to assume relativistic freeze-out, justifying the use of Eq.~\eqref{eq:BBNlifetimelim} as a conservative bound.

\subsection{Applicability of relativistic freeze-out}
The relationship between the Wilson coefficients $c_i$ and the constraint on the HNL lifetime in Eq.~\eqref{eq:BBNlifetimelim} is most straightforward when we can use the simplified lifetime bound in Eq.~\eqref{eq:simplifiedBBNlifetime}.
The simplified bound only holds for HNLs that froze out relativistically ($T_f \gtrsim M_4$), where we can safely assume $Y_N^f \gtrsim 10^{-3}$.
Therefore, we need to verify that $T_f \gtrsim M_4$ when applying the simplified lifetime bound in Eq.~\eqref{eq:simplifiedBBNlifetime}.

To determine the lifetime bound condition, we estimate the HNL decay rates as
\begin{align}
\label{eq:GammaNdecay}
    \Gamma_N^\mathrm{decay} = \sum_i B_i(M_4) |c_i|^2 G_F^2 \frac{M_4^5}{768 \pi^3}\ ,
\end{align}
which is normalized to the decay rate of a Dirac HNL into massless neutrinos (see, e.g., Ref.~\cite{Gorbunov:2007ak, Atre:2009rg}).
The coefficients $c_i$ are Wilson coefficients, and $B_i(M_4)$ accounts for the multiplicity factors of the decay channels, as well as hadronization and threshold effects due to the masses of the final-state particles.
This can cause $B_i$ to deviate from the coefficient in the production rate $A_i$ at lower masses, as shown in Fig.~\ref{fig:alphaePlot2} for a specific example discussed below. 
In the minimal-mixing scenario, we have $|c_0|^2 = 4 \sum_l |U_{l4}|^2$, and for $m_\tau < M_4 < m_b$ we can approximate $B_0(M_4) \approx 24.7$, as it effectively coincides with Eq.~\eqref{eq:AMikko}.

We can now check when both conditions are satisfied simultaneously via the inequality
\begin{align}
    \left. \frac{\Gamma_N^\mathrm{prod}}{H} \right|_{T=M_4} < 1 < \Gamma_N^\mathrm{decay} \tau_\mathrm{BBN}\ ,
\end{align}
where the first part of the inequality imposes relativistic freeze-out and the second the simplified lifetime bound in Eq.~\eqref{eq:simplifiedBBNlifetime}.

Together, these two conditions provide a lower bound on $M_4$, above which the lifetime $\tau_\mathrm{BBN}$ can safely be used to impose upper bounds on the scale of NP.
We obtain
\begin{align}\label{eq:m4bound}
    M_4 > 84.5\, \sqrt{\frac{M_\mathrm{Pl}^\star}{\tau_\mathrm{BBN}} \frac{\sum_i A_i(T_f=M_4)}{\sum_i B_i(M_4)}} \gtrsim 0.44\,\mathrm{GeV} \,\alpha_i(M_4)\ ,
\end{align}
where we have defined $\alpha_i(M_4)\equiv\sqrt{\sum_i A_i(T_f=M_4)/\sum_i B_i(M_4)}$.
The dependence on $c_i$ explicitly cancels, since we are considering a single NP scale.
This cancellation does not occur when multiple NP scales are present.
In the minimal scenario, the two coefficients approximately coincide with $\alpha_0 \approx 1$, which leads to the bound $M_4 \gtrsim 0.44\,$GeV.
In more general $\nu$SMEFT scenarios, we also expect $\alpha_i = \mathcal{O}(1)$ as opening a production channel necessarily implies its presence in decays for sizable HNL masses.
Hence, $A_i$ and $B_i$ are expected to be of the same order of magnitude.
However, there are two possible caveats. 
For HNLs with masses below the GeV scale, production may occur above the QCD confinement scale and involve free quarks. 
In contrast, their decays proceed only through mesonic final states, which modifies the effective values of $\alpha_i$. 
In addition, such HNLs could couple to heavy SM fermions in production at high temperatures, while the corresponding decay channels become kinematically inaccessible at low temperatures, leading to an additional discrepancy between $A_i$ and $B_i$. 

This raises the question of how to apply BBN bounds for HNL masses below the bound in Eq.~\eqref{eq:m4bound}, when the simplified formula cannot be used.
In such a case, one instead has to compute the freeze-out abundance explicitly by solving the Boltzmann equation in Eq.~\eqref{eq:sterile_production} to capture the corresponding Boltzmann suppression.
The resulting abundance should directly enter the computation of the BBN bounds from EM and leptonic decays, which requires a dedicated analysis such as in, e.g., Refs.~\cite{Sabti:2020yrt,Brdar:2020quo}.
As we show in the following subsection, for operators that primarily couple to quarks, even for masses below $100$ MeV, one can assume a relativistic freeze-out as the HNLs freeze out close to the QCD crossover.

\subsubsection{An explicit example}
\label{sub:example}
In a specific EFT scenario, we can more carefully investigate the freeze-out HNL abundances and determine the value of $\alpha_i$ in Eq.~\eqref{eq:m4bound}.
For simplicity, we consider a single non-zero dimension-6 $\nu$SMEFT operator and consider only first-generation SM fermions: $C_{du\nu \ell,111}=1/\Lambda^2$ ($c_{{\rm VRR},111}^{\rm CC}=v^2/\Lambda^2$ after matching to the $\nu$LEFT).

The factor $A_{du\nu \ell,111}$ that appears in Eq.~\eqref{eq:GammaProd} can be determined directly from Eq.~\eqref{eq:AMikko} by considering only the charged-current (CC) contributions of first-generation quarks, and further setting $V_{ud}=1$. This results in $A_{du\nu \ell,111}=2N_c=6\,$.
To determine $B_{du\nu \ell,111}(M_4)$, we calculate the total decay rate $\Gamma_N$ following Ref.~\cite{Gunther:2023vmz}.
For HNL masses $M_4 < 1.2\,$GeV, we compute the decay rate to explicit mesonic final states.
In this mass regime, the decay rate is dominated by decays into a single pion or rho meson (i.e. $\Gamma_N=\Gamma_{N\to e^-+\pi^+}+\Gamma_{N\to e^-+\rho^+}$).
Above it, multi-meson final states become significant.  

\begin{figure}
    \centering
    \includegraphics[width=0.65\linewidth]{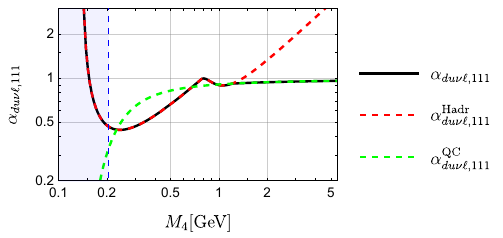}
    \caption{The factor $\alpha_i(M_4)$ as defined in Eq.~\eqref{eq:m4bound} in an EFT scenario where $C_{du\nu \ell,111}=1/\Lambda^2$. The total decay rate $\Gamma_N$ is calculated assuming hadronic substructure (the quark-current approximation) for $\alpha^\text{Hadr}_{du\nu \ell,111}$ $(\alpha^\text{QC}_{du\nu \ell,111})$. In this scenario, the mass threshold for the quark-current approximation is $M_4\gtrsim1.2\,$GeV. The HNLs decouple non-relativistically for masses highlighted in blue.}
    \label{fig:alphaePlot2}
\end{figure}

We approximate contributions of multi-meson decay channels by multiplying HNL decay rates into free-quark currents by appropriate loop corrections through \cite {Bondarenko:2018ptm, DeVries:2020jbs}   
\begin{align}
\label{eq:qcapprox}
    \Gamma_{N\to \ell^-/\nu_\ell +\text{hadrons}}\simeq [1+\Delta_{\rm QCD}(M_4)]\Gamma^{{\rm tree}}_{N\to \ell^-/\nu_{\ell}+\bar{q}_1q_2}\ ,
\end{align}
where
\begin{align}
\label{eq:qcapprox2}
\Delta_{\rm QCD}(M_4)=\sum_{n=1}^3 c_n \left(\frac{\alpha_s(M_4)}{\pi}\right)^n\ ,
\end{align}
with $(c_1,c_2,c_3)=(1,5.5,26.4)$. 
This technique was first introduced to calculate multi-meson corrections to inclusive hadronic decay rates of $\tau$ leptons \cite{Perl:1991gd, Braaten:1991qm, Gorishnii:1990vf}.
We show the value of $\alpha_{du\nu\ell,111}$ as a function of the HNL mass in Fig.~\ref{fig:alphaePlot2}.  

\begin{figure}[h!]
    \centering
    \includegraphics[width=0.9\textwidth]{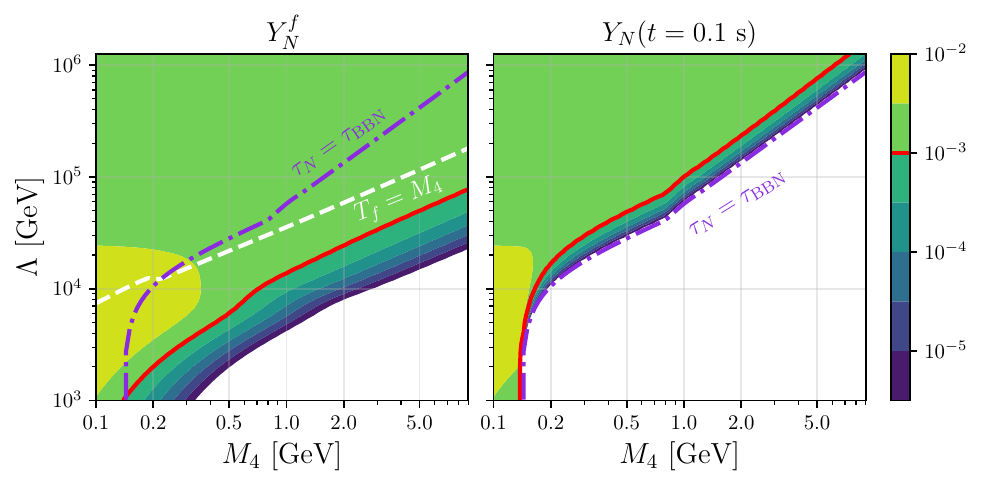}
    \caption{The left panel shows freeze-out HNL abundances $Y_N^f$ obtained by solving the Boltzmann equation in Eq.~\eqref{eq:sterile_production} neglecting HNL decays, while the right panel shows the HNL yield $Y_N$ at the onset of BBN including decays.
    The white dashed line shows the conservative relativistic freeze-out condition from Eq.~\eqref{eq:TfLambda}.
    The simplified BBN bound ($\tau_N > 0.02\,\mathrm{s}$) is shown by the violet dot-dashed line, and is fully compatible with $Y_N^f > 10^{-3}$ in the region of parameter space indicated by the solid red line.
    For smaller HNL masses, we can see that the dot-dashed violet crosses the red line around $M_\pi$, where the HNLs become stable.
    The computation with the HNL decays included (right) shows almost perfect agreement with the simplified violet dot-dashed BBN bound.
    }
    \label{fig:freeze-out}
\end{figure}

For large HNL masses, $\alpha_{du\nu \ell,111}$ approaches unity up to small QCD loop corrections. 
This is unsurprising as both $A_i$ and $B_i$ are essentially determined by multiplicities in HNL interactions and decay, which are closely related for large masses. 
For smaller masses, $\alpha_{du\nu \ell,111}$ deviates from unity because the final-state decay products now involve mesonic states. 
Solving Eq.~\eqref{eq:m4bound} numerically yields a lower mass bound of $M_4>0.21\,$GeV, above which HNLs decouple relativistically. 
However, as illustrated by the freeze-out abundances in the left panel of Fig.~\ref{fig:freeze-out}, the validity of the lifetime bounds imposed by the BBN constraints can be extended somewhat below this value.

To compute the freeze-out abundance $Y_N^f$, we only include the production processes when generating the left panel of Fig.~\ref{fig:freeze-out}, neglecting the contribution from decays that eventually deplete the HNL density.
For convenience, we extrapolate the HNL production rate $\Gamma_N^\mathrm{prod}$ beyond its strict limit of applicability to $T < M_4$.
Furthermore, since we are considering a scenario in which HNLs primarily couple to quarks, at temperatures below the QCD crossover $\Lambda_\mathrm{QCD}$, HNL production is dominated by pion decays.
In this temperature range, the pion density itself becomes Boltzmann suppressed, which we model by multiplying the production rate by a suppression factor $\sim \exp(-m_\pi^2/T^2)$ that appears in the rate for the thermal $\pi^\pm \rightarrow N \ell^\pm$ HNL production process.

We confront the resulting freeze-out abundance with the simplified HNL lifetime bound and the conservative relativistic freeze-out condition.
The two indeed intersect around $M_4 = 0.21\,$GeV, yet the regime where $Y_N > 10^{-3}$ is compatible with the lifetime bound extends to even smaller masses.
There are three reasons for this:
\begin{enumerate}[noitemsep]
    \item The Boltzmann suppression of the HNL number density is only substantial when $T_f < 10\,M_4$
    \item For low enough temperatures, freeze-out can occur in an already diluted plasma with fewer degrees of freedom ($g_\star \sim 10.75$), leading to a larger overall HNL density-to-entropy ratio.
    This enhancement only goes so far and is eventually superseded by the Boltzmann suppression.
    The lifetime bound in Eq.~\eqref{eq:BBNlifetimelim} can then no longer be used without adjusting the value of $Y_N^f$.
    \item For operators that couple to quarks, the HNL production rates quickly become Boltzmann suppressed below the QCD crossover, as the meson masses exceed the temperature of the plasma. 
    This effectively stops the HNL production around $T \sim \Lambda_{\rm QCD}$, acting as a separate freeze-out scale that guarantees relativistic freeze-out for $M_4 \lesssim M_\pi$.
\end{enumerate}

To ensure the validity of the simplified BBN bound, we focus on HNLs with masses above $100\,$MeV in the remainder of this work.
This choice also avoids theoretical uncertainties associated with the QCD crossover.
The main conclusion here is that the BBN bounds remain valid for slightly smaller masses due to plasma dilution, and Eq.~\eqref{eq:BBNlifetimelim} can be safely employed for HNLs with $0.21\,\mathrm{GeV}> M_4> 0.1\,$GeV.

\begin{figure}[t!]
    \centering
    \includegraphics[width=0.45\linewidth]{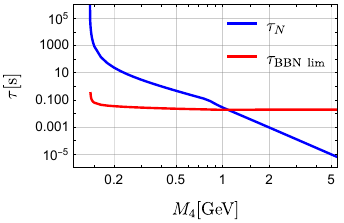}\hspace{0.05\linewidth}\includegraphics[width=0.45\linewidth]{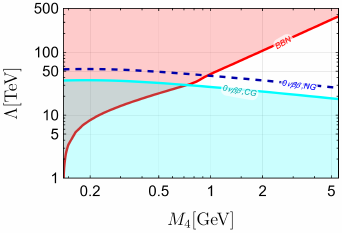}
    \caption{For $C_{du\nu \ell,111}=1/\Lambda^2$ scenario with a fixed $\Lambda=50$~TeV, the left panel shows the HNL lifetime (blue) and the BBN lifetime bound determined by Eq.~\eqref{eq:BBNlifetimelim} (red). 
    The right panel shows the exclusion limits on the NP scale imposed by BBN lifetime constraints (red) and current-generation constraints on the ${}^{136}\text{Xe}$ \ovbb half-life (cyan), along with projected next-generation \ovbb constraints (blue).}
    \label{fig:Cdunue1141Plot2}
\end{figure}

In Fig.~\ref{fig:Cdunue1141Plot2} (left), the total HNL lifetime and the BBN-lifetime bound in Eq.~\eqref{eq:BBNlifetimelim} are shown in a scenario where  $C_{du\nu \ell,111}=1/\Lambda^2$, with the NP scale set to $\Lambda=50\,$TeV. 
In this example, HNLs with masses $M_4<1.2\,$GeV are excluded by BBN constraints.
In Fig.~\ref{fig:Cdunue1141Plot2} (right), such limits are translated into \textit{lower} bounds on the NP as a function of the HNL mass. 
These BBN constraints are complementary to those from neutrinoless double-beta decay and displaced-vertex searches. 
The \ovbb bounds are also included in Fig.~\ref{fig:Cdunue1141Plot2} and are elaborated on in the following section. 
In Sec.~\ref{sec:Results}, we provide similar bounds for other theoretical scenarios. 

We end this section by briefly discussing the situation in which the EFT operator couples to different generations of quarks and leptons.
For example, turning on $C_{du\nu \ell,211}$ and nothing else would lead to a stable HNL with mass below the kaon mass, which rules out the mass range $M_4 < M_K$.
This situation mirrors the one with pions, where HNLs below the pion mass are essentially excluded by BBN bounds as shown in Fig.~\ref{fig:freeze-out}.
On the other hand, the situation becomes more nuanced when multiple operators are present simultaneously.
In particular, in the case where the HNLs couple to leptons, the thermal HNL production does not stop at the QCD scale, but instead continues at lower temperatures, and may even lead to a non-relativistic freeze-out,
requiring a careful treatment of the BBN bounds for $M_4 \lesssim 100\,\mathrm{MeV}$.

\section{Neutrinoless double beta decay}
\label{sec:ovbb}
We consider relatively light Majorana HNLs with non-minimal interactions described by the $\nu$SMEFT Lagrangian. 
In this setup, HNL interactions can yield sizable lepton-number-violating contributions, particularly to neutrinoless double beta decay. 
A general $\nu$SMEFT framework for neutrinoless double beta decay has been developed in Ref.~\cite{Dekens:2020ttz}, and the expressions derived there can be directly applied here.

We give one explicit example for the right-handed charged-current operator in Eq.~\eqref{eq:nuleftCC}. 
The $0\nu\beta\beta$ lifetime can be written as
\begin{equation}
\left(T^{0\nu}_{1/2}\right)^{-1} = g_A^4 G_{01} |\mathcal A_{R}|^2\ ,
\end{equation}
where $g_A \simeq 1.27$ is the nucleon axial charge and $G_{01}$ is the isotope-dependent phase space factor. 
For ${ }^{136}$Xe, which gives the most stringent constraints, we have $G_{01} = 1.5\cdot 10^{-14}$ yr$^{-1}$. 
The amplitude $\mathcal A_{R} = \sum_{i=1}^N \mathcal A_{R}(m_i)$\footnote{For notational convenience, we denote the HNL masses as (lowercase) $m_i$ for $i>3$ in this section, as opposed to the usual $M_i$.} sums the contributions from all neutrino mass eigenstates. 
However, active neutrino contributions are typically suppressed by small mixing angles for the $\nu$SMEFT operators considered in this work. 
Each contribution can be written in terms of a nuclear matrix element $\mathcal M(m_i)$ that depends on the mass of the exchanged neutrino as 
\begin{equation}
\mathcal A_R(m_i) = - \frac{m_i}{4 m_e} \mathcal M(m_i) \left(C_{\rm VRR}^{(6)}\right)^2_{ud ei}\ .
\end{equation}
The presence of the electron mass $m_e$ is purely a conventional normalization, and there is a compensating factor $m_e$ in the definition of the phase space factor $G_{01}$. 
For neutrino masses that are small compared to nuclear scales $\sim\,m_\pi(\simeq 100\,$MeV$)$, the nuclear matrix element (NME) is constant and the amplitude scales as $(m_i/m_e) (v/\Lambda)^4$. 
For large masses, the NME instead scales as $m_\pi^2/m_i^2$, yielding an amplitude that scales as $(m_\pi^2/m_e m_i) (v/\Lambda)^4$. 
At intermediate masses, the scaling behavior is more complicated and has been calculated with a combination of chiral EFT, lattice QCD, and phenomenological techniques \cite{Dekens:2020ttz, Tuo:2022hft, Cirigliano:2024ccq}. 

Similar expressions hold for the other charged-current interactions in Eq.~\eqref{eq:nuleftCC}. 
For the various $\nu$SMEFT operators, we do not provide the lengthy expressions, but instead refer to Refs.~\cite{Dekens:2020ttz, Dekens:2024hlz} for more details. 
In general, the transition from $\sim m_i$ scaling for small masses to $\sim 1/m_i$ scaling for large masses leads to a peak in the $0\nu\beta\beta$ sensitivity for masses $\mathcal O(100)$ MeV. 
The exact location of the peak depends on the specific EFT scenario. 
In the large-mass regime, the inverse \ovbb lifetime scales as $(m_i \Lambda^4)^{-2}$, implying that the corresponding constraints on the EFT scale $\Lambda$ diminish slowly, scaling as $m_i^{-1/4}$.

Using these rough estimates, we can already estimate the sensitivity of $0\nu\beta\beta$ to the EFT operators and the EFT scale. 
We use the existing limit of $T^{0\nu,\mathrm{lim}}_{1/2} = 3.8\cdot 10^{26}\,$y for ${ }^{136}$Xe~\cite{KamLAND-Zen:2024eml}. 
We also consider a prospective improved sensitivity from tonne-scale experiments of $T^{0\nu}_{1/2} > 1.0\cdot 10^{28}\,$y~\cite{nEXO:2017nam}. 
For the existing limit and $m_i \gg m_\pi$, we obtain
\begin{equation}
\Lambda \gtrsim \left( G_{01} T^{0\nu,\mathrm{lim}}_{1/2} \right)^{1/8} \left(\frac{g_A^2 v^4 m_\pi^2}{4 m_e m_i}\right)^{1/4}\simeq (20\,\mathrm{TeV})\times \left(\frac{\mathrm{GeV}}{m_i}\right)^{1/4}.
\end{equation} 
This is just an estimate of the sensitivity, and the numerical value on the rhs changes for other EFT operators and for lighter HNL masses. For our figures, we use the detailed formulae from Ref.~\cite{Dekens:2020ttz}.

\section{Displaced-vertex searches}\label{sec:DV}

HNLs can be produced at the interaction points of high-energy particle colliders. 
For sufficiently long lifetimes, their decays can be observed as reconstructible displaced vertices in dedicated experiments. 
Examples include the recently approved beam-dump experiment \texttt{SHiP} \cite{SHiP:2015vad, Alekhin:2015byh, SHiP:2018xqw, SHiP:2021nfo}, the upcoming long-baseline neutrino facility \texttt{DUNE} \cite{DUNE:2016evb, DUNE:2016rla,DUNE:2020lwj, DUNE:2020ypp, DUNE:2020jqi, DUNE:2021cuw, DUNE:2021mtg} at the \texttt{LBNF}, as well as \texttt{ANUBIS} \cite{Bauer:2019vqk}, \texttt{CODEX-b} \cite{Gligorov:2017nwh}, \texttt{FACET} \cite{Cerci:2021nlb}, \texttt{FASER(2)} \cite{Feng:2017uoz,FASER:2018eoc}, \texttt{MoEDAL-MAPP1(2)} \cite{Pinfold:2019nqj,Pinfold:2019zwp}, and \texttt{MATHUSLA} \cite{Chou:2016lxi,Curtin:2018mvb,MATHUSLA:2020uve}. 
These experiments provide highly sensitive probes for HNLs in the GeV mass range. 

We focus on HNL production through leptonic and semi-leptonic decays of pions, kaons, D mesons, and B mesons. Contributions from heavy pseudoscalar mesons ($B_c$, $J/\Psi$, $\Upsilon$) and other very short-lived mesons ($\eta^{(\prime)}$, vector mesons) are subdominant and therefore neglected. 
For the subsequent HNL decay, both vector mesons $(\rho$, $\omega$, $\phi$, $K^*$,$D_{(s)}^*)$ and pseudoscalar mesons $(\pi, K, \eta^{(\prime)}, D_{(s)}, \eta_c)$ are considered as final-state particles. 
For HNLs below the $B$-meson mass, the production of GeV-scale HNLs via Higgs, $W$- and $Z$-boson decays is subdominant due to relatively smaller production cross-sections. 
The computational details of the relevant HNL production and decay modes in the $\nu$SMEFT framework, as well as the resulting sensitivity reaches for a broad range of experiments, can be found in Refs.~\cite{DeVries:2020jbs,Gunther:2023vmz,deVries:2024mla}. 

For HNLs with masses above the $B$-meson threshold, production through meson decay becomes negligible, and prompt production at interaction points becomes relevant. 
In Ref.~\cite{Beltran:2021hpq}, it was shown that the high-luminosity (HL) phase of the \texttt{LHC} can reach stringent sensitivities to $\nu$SMEFT couplings that induce HNL interactions for masses up to $\mathcal O(50\,\mathrm{GeV})$. 
We employ these sensitivity projections for the \texttt{HL-LHC}, assuming $3\,\mathrm{ab}^{-1}$ integrated luminosity. 
To further compare our results to existing collider bounds, we recast constraints on the minimal-mixing angles $|U_{\ell4}|^2$ provided by the \texttt{BEBC} \cite{WA66:1985mfx,Barouki:2022bkt},  \texttt{E949} \cite{E949:2014gsn}, \texttt{NA62} \cite{NA62:2020mcv}, \texttt{NuTeV} \cite{NuTeV:1999kej}, \texttt{PIENU} \cite{PIENU:2019usb}, \texttt{PS191} \cite{Bernardi:1987ek},  \texttt{SuperK} \cite{Coloma:2019htx}, and \texttt{T2K} \cite{T2K:2019jwa} experiments in terms of the $\nu$SMEFT parameter space \cite{Beltran:2023nli,deVries:2024mla}.

The collider and fixed-target analysis we perform broadly follows Ref.~\cite{deVries:2024mla}. 
However, the expected number of observed HNLs $N_{N}^{\rm obs}$ is calculated slightly differently. 
In previous works, this quantity was obtained with the event generator \texttt{PYTHIA8} \cite{Sjostrand:2014zea, Bierlich:2022pfr} via
\begin{equation}
    N_N^{\mathrm{obs}}=\mathrm{Br}\big(N\rightarrow \mathrm{visible}\big)\sum_M N_{M}\cdot \text{Br}(M\to N+X) \cdot \big<P_M\big[N\text{ }\mathrm{in}\text{ }\mathrm{f.v.}\big]\big>\ ,
\end{equation} 
where Br$(N\to\text{visible})$ denotes the HNL branching ratio into final states for which the detector setup is assumed to be capable of reconstructing the vertex, and a $100\%$ detection efficiency is assumed. 
The sum runs over initial-state mesons $M$, with $N_M$ the total number of mesons $M$ produced during the experiment’s runtime and Br$(M\to N+X)$ the corresponding branching ratio into final states that include an HNL. 
The factor $\big<P_M\big[N\text{ }\mathrm{in}\text{ }\mathrm{f.v.}\big]\big>$ represents the average probability for an HNL to decay within the fiducial volume of a detector.
The decay vertex of $M$ is used to veto long-lived mesons such as kaons that could travel macroscopic distances. 
If such mesons interact with the infrastructure surrounding the interaction points before decaying into an HNL, they do not contribute positively to the average probability. 
In contrast to previous works, we sum not only over initial-state mesons but also over final-states. 
That is, we take $\sum_M\to\sum_{M,X}$ and $P_M\to P_{M,X}$.  

The sensitivity reach of future DV experiments is predicted to provide stringent bounds across the HNL parameter space.
In the UV-motivated EFT scenarios discussed in the following section, we include sensitivity reaches at $95\%\,$C.L.~for three-signal events assuming no background. 
We consider the \texttt{ANUBIS}, \texttt{DUNE}, and \texttt{SHiP} experiments, as these are projected to dominate the DV search sensitivity; cf. Refs.~\cite{DeVries:2020jbs,Gunther:2023vmz,deVries:2024mla} for further details.

\section{Results}
\label{sec:Results}
To avoid providing a lengthy and exhaustive analysis of the extensive $\nu$SMEFT operator basis, we instead consider three representative classes of theoretical scenarios.
In all scenarios, we focus on HNLs that are both long-lived and kinematically relevant for DV searches, such that DV sensitivities can be directly compared with BBN and $0\nu\beta\beta$ constraints. 
We consider simplified UV-motivated EFT scenarios in which we turn on specific $\nu$SMEFT/$\nu$LEFT operators, mimicking the structure of common BSM models.

DV searches typically require at least two non-zero Wilson coefficients (WCs) in order to achieve competitive sensitivity. 
One WC primarily controls HNL production, while the other primarily governs decay. 
In general, the couplings to heavier (lighter) quark states will dominate HNL production (decay).
Accordingly, we consider scenarios with two non-zero WCs, $c^{ijk}$, with up- and down-quark generation indices $i,j$, and a lepton-generation index $k$. 
In all cases, we include a coupling to first-generation quarks, $c^{ud\ell}$, which predominantly induces HNL decays. 
In addition, we turn on a second WC that couples HNLs to heavier quark states and drives their production.

We define three flavor-benchmark scenarios (FBs), corresponding to different dominant production modes: HNL production via $D$-meson decay (FB1), kaon decay (FB2), and $B$-meson decay (FB3). 
These correspond to non-zero values of $c^{cd\ell}$, $c^{us\ell}$, and $c^{ub\ell}$ in FB1, FB2, and FB3, respectively, and are denoted \textbf{FB$\mathbf{1\ell}$}, \textbf{FB$\mathbf{2\ell}$}, and \textbf{FB$\mathbf{3\ell}$}.
For relatively light HNLs, interactions induced by $c^{ud\ell}$ dominate the BBN constraints. For $\ell = e$, the neutrinoless double beta decay rates are solely determined by $c^{ud\ell}$.

\subsection{Right-handed interactions}

\begin{figure}[t!]
    \centering
    \begin{subfigure}{0.3\textwidth}
        \centering
        \includegraphics[width=0.9\textwidth]{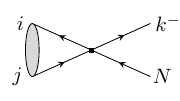}
        \caption{CC two-body production}
        \label{subfig:Prod2bCC}
    \end{subfigure}
    \begin{subfigure}{0.3\textwidth}
        \centering
        \includegraphics[width=\textwidth]{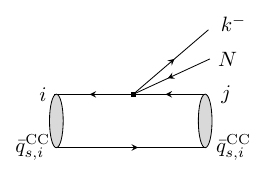}
        \caption{CC three-body production}
        \label{subfig:Prod3bCC}
    \end{subfigure}
        \begin{subfigure}{0.3\textwidth}
        \centering
        \includegraphics[width=0.9\textwidth]{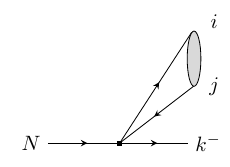}
        \caption{CC decay}
        \label{subfig:Dec2bCC}
    \end{subfigure}
    \begin{subfigure}{0.3\textwidth}
        \centering
        \includegraphics[width=0.9\textwidth]{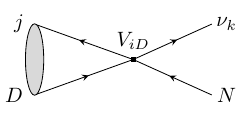}
        \caption{NC two-body production}
        \label{subfig:Prod2bNC}
    \end{subfigure}
    \begin{subfigure}{0.3\textwidth}
        \centering
        \includegraphics[width=\textwidth]{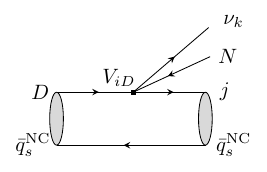}
        \caption{NC three-body production}
        \label{subfig:Prod3bNC}
    \end{subfigure}
        \begin{subfigure}{0.3\textwidth}
        \centering
        \includegraphics[width=0.9\textwidth]{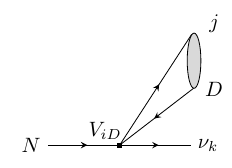}
        \caption{NC decay}
        \label{subfig:Dec2bNC}
    \end{subfigure}
\caption{HNL interaction induced by the WCs $c^{ijk}$ in RH-current scenarios (CC diagrams) and LQ scenarios (CC and NC diagrams). 
Down-type quarks are denoted by $D\in d,s,b$. 
Spectator quarks are $q_s^{\rm NC}\in d,s$ and $q_{s,i}^{\rm CC}=u$. 
In CC three-body production, diagrams with the interchange $i\leftrightarrow j$ also contribute, where $q_{s,j}^{\rm CC}\in d,s$. 
The NC modes are suppressed by the CKM matrix element $V_{iD}$.}
    \label{fig:CPfeynman}
\end{figure}

We first consider RH interactions, inspired by the minimal Left-Right Symmetric Model (mLRSM)~\cite{Mohapatra:1979ia, Pati:1974yy, Mohapatra:1974gc, Senjanovic:1975rk, Mohapatra:1980yp, Senjanovic:1978ev}. 
In this framework, RH neutrinos are charged under a $ SU(2)_R$ gauge symmetry, leading to new gauge fields that give rise to RH heavy gauge bosons $ W_R^\pm $ and $ Z' $. 
After EWSB, these bosons can mix with the SM weak gauge bosons. 
For now, we set the left-right vector-like interactions $c_{{\rm VLR}1}^{\rm CC}=v^2/\Lambda^2$; cf.~Eq.~\eqref{eq:nuleftCC}. We discuss a more complete scenario at the end of this section. 
In the current scenario, HNLs only interact with SM particles through charged-current (CC) interactions via the two- and three-body processes shown in Figs.~\labelcref{subfig:Prod2bCC,subfig:Prod3bCC,subfig:Dec2bCC}. 
The mesonic contents of the interactions are summarized in Tab.~\ref{tab:mesonOverview}

\begin{table}[t!]
    \centering
    \renewcommand{\arraystretch}{1.5}
    \begin{tabular}{|c|cc||c|cc|}
    \hline
       $(q_1 q_2)$  & $M^\pm$ & $M^{*\pm}$ & $(q_1 q_2)$ & $M^0$ & $M^{*0}$\\
      \hline
      ($ud$)  & $\pi^\pm$ & $\rho^\pm$ & ($dd$) & $\pi^0,\eta^{(\prime)}$ & $\rho^0,\omega$\\
       ($us$) & $K^\pm$ & $K^{*\pm}$ & ($ds$) & $K_{L/S}$ & $K^{*0}$ \\
      ($ub$) & $B^\pm$ & $-$ & ($db$) & $B^0$ & $-$ \\
      ($cd$) & $D^\pm$ & $D^{*\pm}$ & ($ss$) & $\eta^{(\prime)}$ & $\phi$ \\
      ($cs$) & $D_s^\pm$ & $D_s^{*\pm}$& ($sb$) & $B_s^0$ & $-$ \\
      \hline
    \end{tabular}
    \caption{Overview of pseudoscalar $(M)$ and vector $(M^*)$ mesons from bound quark states $(q_1\bar{q}_2)$ and $(q_2\bar{q}_1)$ included in the UV scenarios considered in this work. 
    Vector mesons and $\eta^{(\prime)}$ mesons only appear as final-state particles due to their short lifetimes.}
    \label{tab:mesonOverview}
\end{table}

\begin{figure}[t!]
    \centering
    \includegraphics[width=0.48\linewidth]{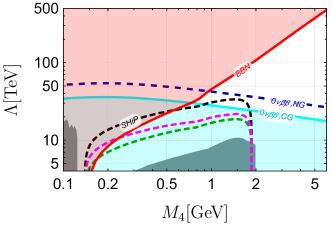}
    \includegraphics[width=0.48\linewidth]{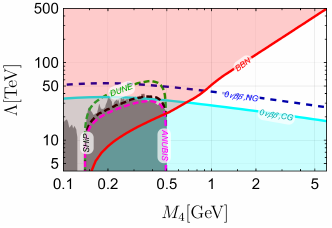}
    \includegraphics[width=0.48\linewidth]{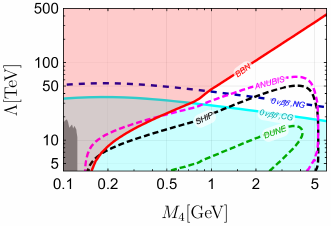}
    \includegraphics[width=0.48\linewidth]{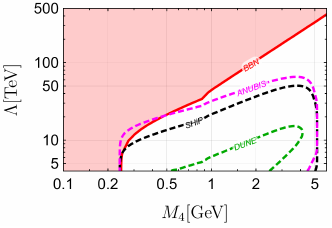}
    \caption{Values of the NP scale $\Lambda$ as a function of the HNL mass for RH-current scenarios. The four panels correspond to \textbf{FB$\mathbf{1e}$} (top left), \textbf{FB$\mathbf{2e}$} (top right), \textbf{FB$\mathbf{3e}$} (bottom left), and \textbf{FB$\mathbf{3\mu}$} (bottom right). 
    Shown are the current exclusion limits from BBN (red), current (cyan) and next-generation (blue) constraints on the ${}^{136}\text{Xe}$ \ovbb half life, the projected sensitivities from DV searches at \texttt{DUNE} (green), \texttt{SHiP} (black) and \texttt{ANUBIS} (magenta), as well as recast limits on the minimal-mixing angle $|U_{\ell 4}|^2$ (gray).}
    \label{fig:LRplot}
\end{figure}

Fig.~\ref{fig:LRplot} shows the current exclusion bounds and future sensitivity reaches on the NP scale $\Lambda$ as a function of the HNL mass for \textbf{FB$\mathbf{1e}$}, \textbf{FB$\mathbf{2e}$}, \textbf{FB$\mathbf{3e}$}, and \textbf{FB$\mathbf{3\mu}$}. 
We show the current exclusion bounds imposed by BBN, and, for the scenarios where $\ell=e$, the \texttt{KamLAND-Zen} limits on the ${}^{136}\text{Xe}$ \ovbb lifetime \cite{KamLAND-Zen:2024eml}. 
We also recast upper bounds on the minimal mixing angles $|U_{\ell 4}|^2$ into lower bounds on $\Lambda$ \cite{Beltran:2023nli}. 
For the future constraints, we illustrate the projected sensitivity of next-generation \ovbb experiments \cite{nEXO:2018ylp}, as well as future DV searches at \texttt{DUNE}, \texttt{SHiP}, and \texttt{ANUBIS}. 

To emphasize the complementarity of the BBN bounds, $0\nu\beta\beta$ isocurves, and DV-search sensitivity reaches, we present results for the HNL mass range $0.1\,$GeV$< M_4 <6.0\,$GeV. 
Although the BBN and $0\nu\beta\beta$ bounds also constrain the NP scale for heavier HNLs, we show this region because the DV searches in our framework are limited to $M_4 \lesssim M_{B_s}=5.37\,$GeV, the heaviest meson we consider for HNL production. 

To verify that the HNL masses are in the range where we can safely use the BBN constraints,
we first numerically solve Eq.~\eqref{eq:m4bound} for $M_4$, where we count a multiplicity factor of $2N_c$ per non-zero WC. 
For this family of operators, we have $A = 4N_c$ for all scenarios.
We find that for \textbf{FB$\mathbf{1e}$}, \textbf{FB$\mathbf{2e}$}, \textbf{FB$\mathbf{3e}$}, and \textbf{FB$\mathbf{3\mu}$},
the HNLs decouple relativistically when the HNL masses exceed $0.28$, $0.21$, $0.28$, and $0.39\,$GeV, respectively. 
For each of the scenarios,
we find that the BBN lifetime bounds can also be safely applied to HNLs with masses below $M_4=0.14\,$GeV,
based on plasma dilution and QCD crossover arguments as discussed in Sec.~\ref{sub:example}.  
For example, for \textbf{FB$\mathbf{1e}$}, the BBN lifetime bounds can be applied to decaying HNLs with masses $M_{\pi^\pm}+m_{e^\mp}=0.14\,$GeV,
while HNLs with $M_4<0.14\,$GeV are excluded as they act as stable relics that would overclose the universe.  

For the scenarios with $\ell=e$, the allowed parameter space broadens at larger HNL masses due to the progressive relaxation of the upper bounds from BBN and the lower bound from \ovbb lower bounds. 
In addition to the \texttt{PIENU} constraints on the minimal-mixing angle $|U_{e4}|^2$, mapped to the NP scale, further constraints are obtained from \texttt{BEBC} and \texttt{CHARM} for \textbf{FB$\mathbf{1e}$}, and from \texttt{NA62} and \texttt{T2K} for \textbf{FB$\mathbf{2e}$}. 
For \textbf{FB$\mathbf{3e}$} and \textbf{FB$\mathbf{2\mu}$}, constraints inferred from existing results, including \texttt{DELPHI} at \texttt{LEP} and recent \texttt{ATLAS} analyses, are not sufficiently restrictive to be shown.

For \textbf{FB$\mathbf{1e}$}, \textbf{FB$\mathbf{2e}$}, and \textbf{FB$\mathbf{3e}$}, the current \ovbb and BBN bounds entirely exclude HNLs with masses below $0.72$, $0.68$, and $0.77\,$GeV, respectively. 
Future \ovbb sensitivities are projected to extend these upper limits to $0.95$, $0.90$, and $0.96\,$GeV. 
For \textbf{FB$\mathbf{2e}$}, this renders the otherwise impressive sensitivity of the \texttt{DUNE} experiment ineffective, since it probes only HNL masses up to the kaon mass, which are already excluded. 
In the remaining allowed mass region, the future \ovbb limits are projected to dominate the lower bounds on $\Lambda$ for \textbf{FB$\mathbf{1e}$} and \textbf{FB$\mathbf{2e}$}. 

For \textbf{FB$\mathbf{3e}$}, in the remaining HNL mass region where the parameter space is not yet closed, \texttt{ANUBIS} is projected to provide the strongest future constraints among DV experiments. 
It outperforms future \ovbb limits for HNL masses $1.36\,$GeV$<M_4<5.22\,$GeV, reaching sensitivities up to $\Lambda=\mathcal{O}(65)\,$TeV around $4\,$GeV. 
At this HNL mass, BBN constraints disfavor scenarios with $\Lambda>\mathcal{O}(250\,\text{TeV})$. 
Achieving sensitivities comparable to this NP scale in terrestrial experiments is a concrete target for future searches aimed at closing the window below the $B$-meson mass. 

In the case of \textbf{FB$\mathbf{3\mu}$}, \ovbb isocurves do not contribute as they vanish in the absence of operators involving first-generation fermion fields ($u$, $d$, $e$, $\nu_e$), or if the HNLs are not Majorana. HNLs are currently excluded for masses $0.1\,$GeV$<M_4<0.27\,$GeV$\simeq m_{\mu^\pm}+m_{\pi^\mp}$, as they would be stable relics. 
The \texttt{ANUBIS} experiment is projected to disfavor HNLs up to $M_4=0.48\,$GeV, attaining sensitivity in scenarios where $\Lambda<\mathcal{O}(70\,\text{TeV})$. 

We note that the DV isocurves shown in Fig.~\ref{fig:LRplot} are quite general across the relevant mass range, as can be noticed comparing \textbf{FB$\mathbf{3e}$} and \textbf{FB$\mathbf{3\mu}$}. Since $m_{e,\mu}/M_B\ll 1$, the isocurves remain largely unchanged when a muon replaces an electron in CC interactions. 
They are furthermore approximately identical for (pseudo-)Dirac neutrinos. 

For \textbf{FB$\mathbf{2e}$} around $M_4 = 0.4\,$GeV, bounds inferred from \texttt{NA62} marginally exceed those from \ovbb within a very narrow mass region. 
Otherwise, \ovbb provides the most stringent lower bounds on $\Lambda$ for all electron-flavor FBs across the HNL mass spectrum. 
The mass ranges where HNLs are already excluded are very similar for all FBs with $\ell=e$. 
In each case, the similarity is driven by the coupling to charged pions induced by $c^{ude}$, which is solely responsible for the \ovbb bounds and dominates the BBN constraints over the other non-zero WCs.

\subsection{Leptoquarks}
The second UV-motivated effective scenario we consider is the leptoquark (LQ) scenario, which introduces hypothetical scalar or vector particles that couple to leptons and quarks, allowing for quark $\leftrightarrow$ lepton transitions. 
We focus on the LQ representation ${\tilde R}_2$, which transforms as $(\boldsymbol{3},\boldsymbol{2},1/6)$ under the SM gauge group. 
For a comprehensive overview of all LQ representations, see Ref.~\cite{Dorsner:2016wpm}. 
The Lagrangian in this scenario is given by 
\begin{align}
    \mathcal{L}^{(6)}_{\rm LQ}=-y_{jk}^{\rm RL}\bar{d}_{Rj}{\tilde R}_2^a\epsilon^{ab}L_{Lk}^b + y_i^{\overline{\rm LR}}\bar{Q}_{Li}^a{\tilde R}_2^a\nu_R + {\rm h.c.}\ ,
\end{align}
with flavor indices $i,j,k$ and $SU(2)$ indices $a,b$. 
Matching to the $\nu$SMEFT Lagrangian after integrating out the heavy LQ gives 
\begin{align}
    C_{LdQ\nu,ijk}^{(6)}=\frac{y_i^{\overline{\rm LR}}{y_{jk}^{\rm LR}}^*}{m_{\rm LQ}^2}\ ,
\end{align}
whereafter matching to the $\nu$LEFT below the EW scale gives 
\begin{align}
    c_{{\rm SRR},ijk}^{\rm CC}=4c_{{\rm T},ijk}^{\rm CC}=\frac{v^2}{2}\frac{y_i^{\overline{\rm LR}}{y_{jk}^{\rm LR}}^*}{m_{\rm LQ}^2},\quad c_{{\rm SRR},ijk}^{\rm NC}=4c_{{\rm T},ijk}^{\rm NC}=-c_{{\rm SRR},ljk}^{\rm CC}V_{li}^* \ .
\end{align}

For the FBs, we set $C_{LdQ\nu}^{(6)}=1/\Lambda^2$, where $\Lambda$ as before is the NP scale, and $\Lambda=m_{\rm LQ}$ if all Yukawa couplings are fixed to unity. 
In this limit, direct LHC searches and indirect flavor and precision constraints each impose TeV-scale lower bounds on scalar leptoquark masses~\cite{ParticleDataGroup:2024cfk}. 
Compared to the RH-current scenario, the HNLs have considerably more interaction modes due to the availability of NC channels. 
For the HNL decay rates, we approximate the multi-meson final-state contributions using the quark-current approximation in Eq.~\eqref{eq:qcapprox}, setting the loop corrections $\Delta_{\rm QCD}(M_4)=0$ since they are not well understood for scalar and tensor currents. 
All relevant HNL interaction modes are shown in Fig.~\ref{fig:CPfeynman}, with the corresponding mesons listed in Tab.~\ref{tab:mesonOverview}. 

\begin{figure}[t!]
    \centering
    \includegraphics[width=0.6\linewidth]{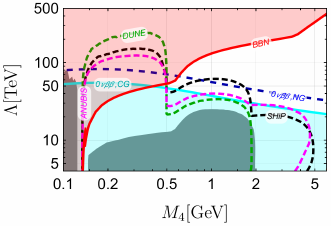}
    \caption{The NP scale $\Lambda$ versus the HNL mass for \textbf{FB$\mathbf{1e}$} in a leptoquark scenario. Included are BBN exclusions (red), current (cyan) and next-generation (blue) constraints on the ${}^{136}\text{Xe}$ \ovbb half-life, projected DV sensitivities from \texttt{DUNE} (green), \texttt{SHiP} (black) and \texttt{ANUBIS} (magenta), as well as recast bounds from \texttt{PIENU}, \texttt{BEBC}, and \texttt{CHARM}.}
    \label{fig:FB1LQplotZoom}
\end{figure}

\begin{figure}
    \centering
    \includegraphics[width=0.495\linewidth]{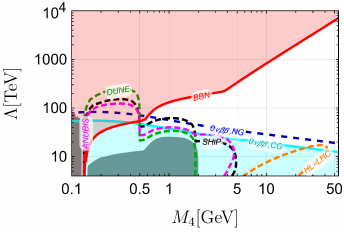}
    \includegraphics[width=0.495\linewidth]{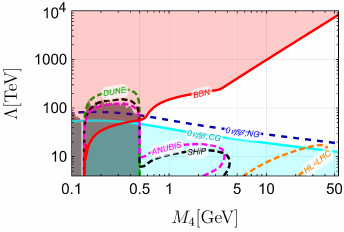}
    \includegraphics[width=0.495\linewidth]{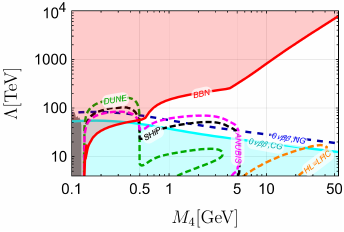}
    \includegraphics[width=0.495\linewidth]{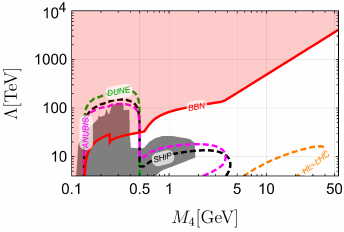}
    \caption{Projected sensitivity reaches and current exclusion bounds on the NP scale $\Lambda$ as a function of the HNL mass for LQ scenarios \textbf{FB$\mathbf{1e}$} (top left), \textbf{FB$\mathbf{2e}$} (top right), \textbf{FB$\mathbf{3e}$} (bottom left), and \textbf{FB$\mathbf{2\mu}$} (bottom right). 
    The solid lines are current exclusion limits from current-gen (CG) \ovbb rates (cyan), BBN (red), and recast existing bounds on the minimal-mixing angle $|U_{\ell 4}|^2$ (gray). 
    Also shown are projected sensitivities from the DV experiments \texttt{DUNE} (green), \texttt{ANUBIS} (magenta), and \texttt{SHiP} (black), as well as recast projected bounds from the \texttt{HL-LHC} \cite{Beltran:2021hpq} (orange).}
    \label{fig:LQplot}
\end{figure}

To again highlight the complementarity of the BBN, \ovbb and projected DV search constraints, we zoom in on the HNL mass range $0.1\,$GeV$<M_4<6.0\,$GeV in Fig.~\ref{fig:FB1LQplotZoom}, where we consider \textbf{FB$\mathbf{1e}$}. 
To determine the relativistic freeze-out limit of HNLs, as before, we count a multiplicity factor of $2N_c$ per non-zero WC and thus find $A=4N_c$ for all FBs. 
Across the entire relevant mass range, we find that HNLs either decouple relativistically or are stable for masses below the threshold $M_4=M_{\pi^0}=135\,$MeV.\footnote{We note that the factor $\alpha(M_4)$ in Eq.~\eqref{eq:m4bound} does not converge exactly to unity in the LQ scenarios considered in this subsection due to RGE effects, with the deviations reaching at most $\mathcal{O}(15\%)$. 
This does not impact our conclusions.} 
At $M_4=M_{\pi^\pm}+m_{e^\mp}=0.14\,$GeV, the BBN curve exhibits a pronounced change in slope as the charged pion modes become available. 
For HNLs below this mass, only neutral decay modes are accessible, and efficient neutron–to-proton conversion ceases.
Consequently, the constraint of Eq.~\eqref{eq:BBNlifetimelim} does not apply, and we impose the conservative bound $\tau_N<0.1\,$s. 
The resulting kink in the BBN curve is more visible for heavier final-state charged leptons; see \textbf{FB$\mathbf{2\mu}$} in Fig.~\ref{fig:LQplot}.

Currently, all HNLs in \textbf{FB$\mathbf{1e}$} with masses $M_4<0.40\,$GeV are excluded through the combined \ovbb and BBN constraints. 
In combination with the BBN bounds, the future \ovbb bounds are projected to further disfavor the HNL parameter space up to $M_4=0.60\,$GeV. For $0.85\,$GeV$<M_4<1.70\,$GeV, \texttt{SHiP} is projected to disfavor NP scenarios up to $\Lambda=\mathcal{O}(60\,\text{TeV})$. 
Outside this window, the future \ovbb constraints will provide the most stringent bounds on the NP scale from below. 

Fig.~\ref{fig:LQplot} shows the experimental sensitivities and exclusion limits for several LQ scenarios across the HNL mass range $0.1\,$GeV$< M_4 < 55\,$GeV for the benchmark configurations \textbf{FB$\mathbf{1e}$}, \textbf{FB$\mathbf{2e}$}, \textbf{FB$\mathbf{3e}$}, and \textbf{FB$\mathbf{2\mu}$}. 
In all scenarios, the HNLs are either stable relics or decouple relativistically, allowing us to employ the same BBN constraints as before. 
We present the same sensitivity reaches and current exclusion bounds shown in Fig.~\ref{fig:LRplot}, together with recast projected bounds from the \texttt{HL-LHC} on the minimal-mixing angle $|U_{\ell 4}|^2$. 
We recast the minimal-mixing-angle bounds using the same experimental inputs as in the previous subsection for \textbf{FB$\mathbf{1e}$}, \textbf{FB$\mathbf{2e}$}, and \textbf{FB$\mathbf{3e}$}. 
For \textbf{FB$\mathbf{2\mu}$}, we show recast bounds on the minimal-mixing angle $|U_{\mu 4}|^2$ from the \texttt{NuTeV}, \texttt{PS191}, \texttt{E949}, \texttt{T2K}, \texttt{SuperK}, and \texttt{NA62} experiments. 

For \textbf{FB$\mathbf{2e}$} (\textbf{FB$\mathbf{3e}$}), HNLs with masses $M_4 < 0.37\,$GeV ($0.50\,$GeV) are disfavored. 
For both \textbf{FB$\mathbf{2e}$} and \textbf{FB$\mathbf{3e}$}, future constraints could further disfavor HNLs with masses up to $0.58\,$GeV. 
For scenarios with $\ell=e$, the \texttt{HL-LHC} sensitivity reaches are outperformed by the future \ovbb constraints. 
For \textbf{FB$\mathbf{2e}$}, the \ovbb constraints outperform all future DV searches for HNL masses above the kaon mass. For \textbf{FB$\mathbf{3e}$}, \texttt{ANUBIS} is projected to be the most stringent in the mass range $0.91\,$GeV$<M_4<5.14\,$GeV, reaching sensitivities of $\Lambda=\mathcal{O}(70\,\text{TeV})$. 

For \textbf{FB$\mathbf{2\mu}$}, the slope change in the BBN curve at the charged-pion threshold ($M_4=M_{\pi^\pm}+m_{\mu^\mp}=0.27\,$GeV) is indeed more pronounced than in $\ell=e$ scenarios. 
The combined lower bounds on $\Lambda$ from the recast bounds on $|U_{\mu 4}|^2$ and the BBN bounds exclude HNLs with masses $M_4 < 0.38\,$GeV. 
The DV searches at \texttt{DUNE}, \texttt{ANUBIS}, and \texttt{SHiP} are projected to further exclude HNLs up to the kaon mass ($M_4=0.5\,$GeV). 
The \texttt{NuTeV} experiment provides the most stringent bounds from below up to $\Lambda=\mathcal{O}(25\,\text{TeV})$ around $M_4=1\,$GeV. 
Around $M_4=3\,$GeV, \texttt{ANUBIS} is projected to disfavor NP scenarios for $\Lambda<\mathcal{O}(18\,\text{TeV})$. 
For larger masses, between $6\,$GeV$< M_4 < 40\,$GeV, the \texttt{HL-LHC} is projected to be sensitive to scenarios reaching $\Lambda=\mathcal{O}(16\,\text{TeV})$.

For the FBs with $\ell=e$, it is no coincidence that the currently excluded HNL masses span a similar range. 
The BBN bounds below the kaon mass are dominated by the HNL decay into charged pions, rendering the differences in neutral pion decay rates between these benchmark scenarios subleading. 
Moreover, the \ovbb constraints are also equal between the scenarios as they are only generated by the operator corresponding to the WC $c^{ude}$. 

Comparing the LR and LQ scenarios, the LQ scenarios exhibit a richer phenomenology due to the numerous NC contributions to the HNL decay rate. 
However, the current exclusion limits and the projected future sensitivity reaches, especially for scenarios with $\ell=e$, are quite comparable since \ovbb limits dominate in a sizable part of the HNL parameter space. Furthermore, above the kaon mass, DV searches are projected to be competitive with the future \ovbb constraints in similar mass regions.

\subsection{mLRSM as a UV completion}
\label{sec:mLRSMUV}
While UV-motivated EFT scenarios provide a practical framework for capturing low-energy signatures of HNLs, they inherently lack the structural coherence of a full theory. 
In particular, it permits unphysical decoupling between production and decay channels and the choice of lepton/quark-flavor structures that can sidestep otherwise stringent experimental constraints. 
To address these limitations and embed HNL phenomenology in a more realistic setting, we assume the mLRSM as the UV completion. 
We follow the detailed mLRSM analysis of DV searches and \ovbb of Ref.~\cite{deVries:2024mla}. 
This will also serve as a useful comparison with the naive operator-by-operator analysis of this model. 

The model details are rather lengthy and are provided in Refs.~\cite{Mohapatra:1979ia, Pati:1974yy, Mohapatra:1974gc, Senjanovic:1975rk, Mohapatra:1980yp, Senjanovic:1978ev,deVries:2024mla}. 
Here, we will focus on the model's main features. 
The extended gauge group of the mLRSM is
\begin{align}
    G_{\rm LR}\in SU(2)_L\otimes SU(2)_R\otimes U(1)_{B-L},
\end{align}
where fermions transform in left- and right-handed $SU(2)$ doublets. 
The $SU(2)_R$ is spontaneously broken at a scale $v_R$, giving rise to massive $W_R$ and $Z^\prime$ gauge bosons. 
In particular, $M_{W_R} = g v_R/\sqrt{2}$ where $g$ is the SM $SU(2)_L$ gauge coupling.

At a lower scale $v \ll v_R$, the gauge groups are broken to $SU(2)_L\otimes U(1)_Y$. 
Neutrino masses are generated via type-I and type-II seesaw mechanisms. 
We will consider the type-II limit, where neutrino masses are generated through another vev, $v_L \ll v$, as this limit leads to the fewest model parameters. 
In this scenario, we obtain a direct relation between active and HNL masses
\begin{align}
\widehat{M}_N=\frac{v_R}{v_L}\widehat{m}_\nu\,,
\end{align}
where $\widehat{M}_N$ and $\widehat{m}_\nu$ are diagonal $3\times 3$ mass matrices of, respectively, sterile and active neutrino masses. 
If we specify the hierarchy (inverted or normal), the lightest neutrino mass $m_1$, and the heaviest HNL $M_6$, all other neutrino masses are fixed. 
Integrating out the $W_R$ gauge bosons leads to $\nu$SMEFT interactions\footnote{Additional terms are generated by integrating out $Z_R$, but they play no significant role in the phenomenology of this work \cite{deVries:2024mla}.}
\begin{align}\label{mLRSMmatch}
    C_{du\nu \ell,ij}^{(6)}=-\frac{V_{R,ij}}{v_R^2},\quad C_{H\nu \ell}^{(6)}=-\frac{1}{v_R^2}\frac{2\xi}{1+\xi^2}e^{-i\alpha}\,,
\end{align}
in terms of the $v_R$, the right-handed analogue of the CKM matrix $V_R$, and the left-right mixing parameter $\xi$. 
Since  $\xi<0.8$ ensures the mLRSM scalar sector remains perturbative \cite{Maiezza:2010ic}, we choose the benchmark values $\xi=0,0.3$. 
We consider the $P$-symmetric mLRSM, in which case the matrix $V_R = V$, where $V$ is the SM CKM matrix, up to small corrections \cite{Senjanovic:2014pva, Senjanovic:2015yea, Dekens:2021bro}. 
Through Eq.~\eqref{mLRSMmatch}, it is clear that in this model all quarks are involved in the BSM interactions. This is distinct from EFT scenarios, in which we consider only one or two Wilson coefficients at a time. 

To calculate for which masses HNLs decouple relativistically, we first have to estimate the multiplicity factor $A_{\rm mLRSM}$ for the HNL production rate that appears in Eq.~\eqref{eq:GammaProd}. 
The contributions from the $C_{du\nu\ell}$ WCs can be read directly from the CC quark contributions in the minimal scenario in Eq.~\eqref{eq:AMikko} since $V_R=V$, and $W_R$ is furthermore completely analogous to $W_L$. 
This implies $A_{du\nu\ell}=A_{\rm CC}^{\rm quarks}=2N_c\sum_{i=u,c}(|V_{id}|^2+|V_{is}|^2+|V_{ib}|^2)=4 N_c$ by CKM unitarity. 

The HNL production channels induced by a non-zero $C_{H\nu\ell}$ for $\ell=e,\mu,\tau$ suggest $A_{H\nu\ell}=|\tfrac{-2\xi}{1+\xi^2}e^{i\alpha}|^2(A_{\rm CC}^{\rm quarks}+A_{\rm CC}^{\rm leptons})$, where $A_{\rm CC}^{\rm leptons}=2\times3$. 
As mentioned before, if a production channel is kinematically allowed, such a channel will likewise appear in the decay of sufficiently heavy HNLs. 
This implies that for a correctly chosen $A_{\rm mLRSM}$, the factor $\alpha(M_4)$ in Eq.~\eqref{eq:m4bound} is expected to converge to unity for large $M_4$. 
Indeed, for $A_{\rm mLRSM}= A_{H\nu\ell}+A_{du\nu\ell}$, we find $\alpha(M_4\to\infty)\to1$ for both benchmark values of $\xi$. For $\xi=0.0$ ($0.3$), relativistic decoupling occurs for HNLs masses $M_4>0.43\,$GeV ($0.26\,$GeV)
In both cases, we may use the BBN lifetime bounds imposed by Eq.~\eqref{eq:BBNlifetimelim} for HNLs down to $0.1\,$GeV employing plasma-dilution arguments similar to those discussed in Sec.~\ref{sub:example}. 

\begin{figure}
    \centering
    \includegraphics[width=0.495\linewidth]{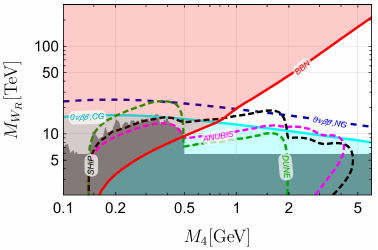}\includegraphics[width=0.495\linewidth]{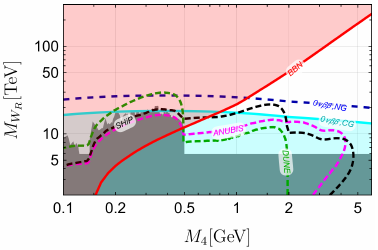}
    \caption{Limits on $M_{W_R}$ in the type-II seesaw scenario without  ($\xi=0$, left panel) and with $W_L-W_R$ mixing  ($\xi=0.3$, right panel).}
    \label{fig:typeIIMWRlims}
\end{figure}

In Fig.~\ref{fig:typeIIMWRlims}, we show constraints on the mass of the RH gauge boson $W_R$ as a function of HNL mass $M_4$. 
We include current bounds imposed by BBN, \ovbb experiments, the minimal-mixing angle $|U_{e4}|^2$ from \texttt{T2K}, \texttt{PIENU}, and \texttt{NA62}, and the latest results from the \texttt{LHC} Run-2 results, which exclude $M_{W_R}<5.7\,$TeV \cite{Li:2025tmt}. 
Regarding future experimental constraints, we show the projected sensitivities of the DV experiments \texttt{SHiP}, \texttt{ANUBIS}, and \texttt{DUNE}, as well as the future \ovbb limits as predicted by \texttt{KamLAND-Zen}.

For $\xi=0.0$ (left panel), the combined BBN and \ovbb constraints currently disfavor all HNLs in the mass range $0.1\,$GeV$<M_4<0.77\,$GeV, and future \ovbb searches could constrain the parameter space further up to HNL masses $M_4=1.0\,$GeV. 
The projected DV sensitivity reaches and the future \ovbb bounds are highly competitive, with both the \texttt{SHiP} experiment and the \ovbb searches probing $W_R$ masses reaching $M_{W_R}=\mathcal{O}(18\,\text{TeV})$ at $M_4= 1.6\,$GeV. 
For $M_4>1\,$GeV, the \texttt{SHiP} constraints slightly dominate the future \ovbb constraints between $1.37\,$GeV$<M_4<1.83\,$GeV. 
Elsewhere in this mass regime, the future \ovbb are projected to provide the most stringent bounds on the RH gauge boson mass from below. 

While completing this work, an mLRSM study of sub-GeV HNLs without left-right mixing in the type-II limit has been performed \cite{Li:2025tmt}, which also discusses BBN bounds and finds results in general agreement with ours. 

In the scenario where the mixing parameter $\xi=0.3$ (right panel), HNLs with masses $0.1\,$GeV$<M_4<0.82\,$GeV are currently excluded, and future \ovbb limits could further disfavor HNLs with masses up to $1.16\,$GeV. 
For heavier HNLs, future \ovbb are predicted to outperform all DV searches. 
As was the case for the LQ scenarios with $\ell=e$, the impressive sensitivity of the \texttt{DUNE} experiment below the kaon mass is ineffective for both mLRSM scenarios considered. 
Differences between the DV sensitivity reaches are caused by cross-terms between the WCs $C_{\rm VRR }^{(6)}$ and $C_{\rm VLR }^{(6)}$, which either suppress or enhance three-body mesonic HNL production rates based on the parity nature of the final-state meson \cite{deVries:2024mla}.

Comparing the simplified RH-current scenario to the full mLRSM picture, we observe similar shapes for kaon, D meson, and B-meson curves, although quark transitions are CKM-suppressed in the mLRSM. 
The similarity implies that the RH-current scenario, or at least a combination of the corresponding flavor benchmarks, is indeed a good approximation of more general mLRSM scenarios. 
This similarity is further reinforced by the fact that DV isocurve shapes remain recognizable even for sizable left–right mixing $\xi$. 
This demonstrates that the $\nu$SMEFT analyses provide a useful framework to describe low-energy interactions of HNLs. 

\section{Conclusions}
\label{sec:conclusions}
Discovery of the elusive heavy neutral lepton (HNL) could provide a solution to a multitude of major open questions in cosmology and particle physics. 
Besides serving as an elegant explanation of the light SM neutrino masses, HNLs also function as prime candidates in further explaining dark matter and the baryon asymmetry of the universe. 

The potential presence of HNLs in the early universe should, however, not interfere with well-established SM light-element abundances during Big Bang Nucleosynthesis (BBN). 
In studies of HNLs that interact with the SM exclusively through minimal active–sterile mixing, the size of the relevant mixing angle solely governs HNL production and decay, and therefore determines the thermalization history. 
In this case, BBN imposes particularly stringent constraints on the HNL lifetime, which can be directly translated into lower bounds on these mixing angles. 

The non-observation of HNLs in terrestrial experiments further imposes stringent upper bounds on these minimal mixing angles. 
Such limits are determined by particle collider searches and lifetime bounds on isotopes that could undergo neutrinoless double beta decay ($0\nu\beta\beta$). 
The combined terrestrial and cosmological constraints in the minimal scenario thus provide clear target regions for future laboratory probes, since the HNL parameter space is bounded both from above and below. 

Beyond the minimal scenario, assuming a sufficient separation between the HNL mass and the scale of new physics, dynamics of relatively light HNLs may instead be described by higher-dimensional operators within the neutrino-extended Standard Model Effective Field Theory ($\nu$SMEFT) above the electroweak scale, and the $\nu$LEFT framework below it. 
In this setting, the link between HNL lifetimes and BBN constraints is not immediately obvious, as many parameters can contribute to HNL production, decay, and, in turn, the thermalization history. 
Moreover, if HNLs freeze out above the EFT validity scale, their thermal history cannot be reliably described by the EFT framework. 
Consequently, whether BBN constraints map straightforwardly onto lifetime bounds in such EFT frameworks is inherently model-dependent. 

Under the relatively minimal assumption of sufficiently high reheating temperatures, we showed that the freeze-out temperatures of EFT operators can be computed without specifying the UV completion for scenarios with a single HNL. 
As such, the $\nu$SMEFT framework can be used for both BBN considerations as well as low-energy laboratory experiments, and BBN constraints provide robust upper bounds on the NP scale for HNLs with masses exceeding $100\,$MeV. 
Lower bounds on the NP scale from terrestrial experiments can be complemented by BBN constraints, providing useful target regions for laboratory HNL probes in more general scenarios than minimal mixing. 

We focused on HNLs with masses ($0.1<M_4<60$)\,GeV, where the BBN bounds on decays into pions are most directly applicable.
For HNL masses below the $B$-meson mass, especially stringent constraints arise from displaced-vertex (DV) searches, since such HNLs can be produced abundantly in pseudoscalar meson decays at collider interaction points.
These constraints typically exceed those from potential direct production via Higgs, $W$-, and $ Z$-boson decays. 

We have performed Monte Carlo simulations to determine the sensitivity reaches of future displaced-vertex (DV) search experiments, considering HNL production from decays of $B$ mesons, $D$ mesons, kaons, and pions. 
We focused on the future DV experiments projected to be most constraining in the HNL parameter spaces under consideration. 
These are the LHC far detector \texttt{ANUBIS}, the \texttt{DUNE} near detector at the \texttt{LBNF}, and CERN's beam-dump experiment \texttt{SHiP}.  
Furthermore, we have provided terrestrial bounds on the NP scale by calculating neutrinoless double-beta-decay rates.

We studied left-right vector-like operator (LR) scenarios and compared them to the more involved UV-complete minimal Left-Right Symmetric Model (mLRSM) in the type-II seesaw dominance limit, where HNLs interact with SM particles through heavy right-handed gauge bosons arising from an $SU(2)_R$ gauge symmetry at high energies. 
We also explored leptoquark (LQ) scenarios that include both charged- and neutral-current interactions via scalar- and tensor-like currents. 

In LR (LQ) scenarios containing operators that generate charged-current interactions between HNLs and first-generation SM fermions, HNLs with masses below $0.7\,$GeV ($0.4\,$GeV) were found to be largely disfavored, with only minor dependence on the choice of additional operators coupling HNLs to quarks.
While the BBN bounds are no longer directly applicable below $\sim30$ MeV,
lighter HNL masses may be subject to other cosmological constraints that go beyond the scope of the current work.
Future \ovbb limits could further extend the disfavored mass ranges, potentially excluding HNLs with masses reaching $\sim0.9\,$GeV ($0.6\,$GeV).
In scenarios where the relevant operators instead couple HNLs to muons — rendering \ovbb bounds inapplicable — the combination of BBN constraints and projected DV sensitivities was shown to close the $\nu$SMEFT parameter space for HNL masses in the interval $(0.1 < M_4 < 0.5)\,$GeV. 

Within the examined class of mLRSM scenarios, the main benchmark parameters governing HNL interactions are the mass of the right-handed charged gauge boson, $M_{W_R}$, and the parameter $\xi$, which quantifies the mixing strength between $W_L$ and $W_R$. 
In a scenario without $W_L-W_R$ mixing ($\xi=0.0$), we find HNLs are currently disfavored between ($0.1<M_4<0.77$)$\,$GeV, with future \ovbb constraints further extending this mass range to HNLs up to $M_4=1.0\,$GeV. 
For a relatively sizable $W_L-W_R$ mixing ($\xi=0.3$), the excluded current upper limit rises to $M_4=0.82\,$GeV, with the projected \ovbb constraints extending the exclusion limit to $M_4=1.16\,$GeV. 

In contrast to the LR (LQ) scenarios, where the HNLs necessarily couple to quarks, in the mLRSM scenarios, HNLs can decay into both hadronic and leptonic final states.
Therefore, HNLs below the pion mass threshold are no longer stable, and the somewhat weaker bounds from leptonic decays need to be used instead.
Interestingly, this happens in a similar mass range where we can no longer assume a purely relativistic freeze-out,
therefore requiring a careful analysis of both the HNL freeze-out and subsequent leptonic/EM decays during BBN.
This caveat is not limited to the mLRSM, but applies more broadly to any scenario that allows for non-hadronic decays,
especially for HNL masses below the pion mass.
This represents a promising direction for future studies of BBN bounds on HNLs, both because of the possibility of a Boltzmann-suppressed freeze-out abundance and because of weaker constraints on leptonic and EM decays.

We conclude that while laboratory experiments are effective at constraining lower bounds on the scale of NP, BBN offers a strong complementary cosmological probe by imposing robust upper bounds in many cases. 
Altogether, the combination of constraints from BBN and terrestrial experiments defines target regions in the $\nu$SMEFT parameter space that provide realistic, well-constrained windows for discovering or excluding GeV-scale HNLs in future studies. 

Extensions of this work could include incorporating partonic processes across all scenarios into future HNL searches at particle colliders, including the \texttt{HL-LHC} and \texttt{FCC-ee}. 
Below the mass threshold for mesonic HNL production, these prompt-production effects are sub-leading, but would improve the accuracy of projected sensitivity reaches for HNLs produced at interaction points. 
For heavier HNLs, it would yield relevant lower bounds on the NP scales.

\section*{Acknowledgment}

JdV, JG, and JK acknowledge support from the Dutch Research Council (NWO) through a VIDI grant. JdV thanks Miguel Escudero regarding clarifications about BBN studies during the initial stages of this work. 

\appendix
\appendix
\section{HNL Production in the Early Universe}
\label{sec:BE_sterile}
The density of HNLs evolves according to the following Boltzmann equation:
\begin{align}
   \dot{n}_N + 3 H n_N = - \Gamma_N (n_N - n_N^\mathrm{eq})\,,
\end{align}
where $n_N$ is the HNL number density, $n_N^\mathrm{eq}$ is the number density in equilibrium, $\Gamma_N$ is the HNL equilibration rate,
and $H$ is the Hubble rate
\begin{align}
    H = \frac{\dot a}{a} = \frac{T^2}{M_\mathrm{Pl}^\star}\,,
\end{align}
assuming radiation domination, the reduced Planck mass is given by:
\begin{align}
    M_\mathrm{Pl}^\star = M_\mathrm{Pl} \sqrt{\frac{45}{4 \pi^3 g_\star(T)}}\,,
\end{align}
and $M_\mathrm{Pl} = 1.2209 \cdot10^{19}$ GeV, with the effective number of relativistic degrees of freedom $g_\star$, that varies between $g_\star \approx 106.75$ in the symmetric phase of the standard model and $g_\star \approx 10.75$ at the beginning of BBN.

The time and temperature are related by~\cite{Laine:2015kra}:
\begin{align}
    \frac{d T}{d t} = - 3 c_s^2 H T \equiv - \mathcal{H} T\ ,
\end{align}
with the sound speed $c_s$, and where we introduced the effective Hubble parameter $\mathcal{H}$ with $\mathcal{H} \approx H$ during most of the thermal history of the universe.

If we normalize the HNL density to the entropy density with
\begin{align}
    Y_N \equiv \frac{n_N}{s}\,,
\end{align}
the HNL rate equation simplifies to
\begin{align}
   \dot{Y}_N = - \Gamma_N (Y_N - Y_N^\mathrm{eq})\ ,
\end{align}
where we used that $s^\prime = s/(T c_s^2)$.
By changing the time variable to $z=M/T$, we obtain the rate equation in Eq.~\eqref{eq:sterile_production}.

The interaction rate $\Gamma_N$ contains both the production and decay channels, and it parametrically scales as:
\begin{align}
	\Gamma_N \sim
	\begin{cases}
		\frac{T^5}{\Lambda^4} \text{ for } T>M\ ,\\
		\frac{M^5}{\Lambda^4} \text{ for } M>T\ ,\\
	\end{cases}
\end{align}
with the latter case corresponding to decays.
The HNL production effectively stops when $\Gamma_N/H < 1$. In the relativistic limit, this sets the freeze-out temperature:
\begin{align}
    \label{def:Temp_freeze-out}
    T_f = \Lambda \sqrt[3]{\frac{\Lambda}{M_\mathrm{Pl}^*}}\ ,
\end{align}
where $M_\mathrm{Pl}^\star$ follows the usual definition
\begin{align}
    M_\mathrm{Pl}^\star \equiv M_\mathrm{Pl} \sqrt{\frac{90}{8 \pi^3 g_\star}}\ ,
\end{align}
and $M_\mathrm{Pl} = G^{-1/2}$ is the Planck mass, and $g_\star$ is the effective number of degrees of freedom, with $g_\star \approx 106.75$ for the SM fully in equilibrium.

In the case where $T_f > M_4$, the HNLs freeze out while relativistic, and their yield is set by:
\begin{align}
    Y_N^f \approx Y_N^\mathrm{eq}(T = T_f) \approx \frac{3 \zeta(3)}{4 \pi^2 s(T_f)} \approx 1.95 \cdot 10^{-3}\ ,
\end{align}
where we assumed that the SM was fully in equilibrium during freeze-out at the last step.
This also sets the HNL energy density:
\begin{align}
    \rho_N \approx Y_N^f s(T) \cdot \langle E_N \rangle\ ,
\end{align}
with $\langle E_N \rangle$ as the average HNL energy, which approaches $E_N \approx M_4$ in the non-relativistic limit.
The HNL abundance then remains fixed until $\Gamma_N/H > 1$ again due to the HNL decays,
this corresponds to $t \approx \tau_N = \Gamma_N^{-1}$.

\section{HNL production rate}
The vector channel of HNL production below the electroweak crossover can be directly interpreted from the minimal scenario (see, e.g.,~\cite {Asaka:2006nq,Ghiglieri:2016xye}).
For illustrative purposes, we show how the production can be computed for scalar four-fermion operators.
The production rate is given by the imaginary (spectral) part of the retarded HNL self-energy:
\begin{align}
    \Gamma(k) = \frac{\Tr (\slashed{k} \slashed{\Sigma}_N^\rho)}{2 k_0}\,.
\end{align}
The self-energy can be expressed as:
\begin{align}
    i \Sigma_N^{ab} =
    2 G_F^2 |c_{S,ijk}|^2 A_{S,ijk} \int \frac{d^4 q}{(2 \pi)^4} \frac{d^4 p}{(2 \pi)^4}
    i S_k^{ab}(k-q) \Tr \left[ i S_j^{ab}(p) i S_i^{ba}(p+q) \right]\,,
\end{align}
where we suppressed the CC, NC, and L, R indices in $c_{S,ijk}$. For simplicity, we also assume that $i\neq j\neq k$.
The two momentum integrals factorize, and after using the equilibrium propagators, we can express the self-energy as
\begin{align}
\Sigma_N^\rho = 8 G_F^2 |c_{S,ijk}|^2 A_{S,ijk} & \int  \frac{d^4 q}{(2 \pi)^4} \frac{d^4 p}{(2 \pi)^4}
    S_k^{\rho}(k-q) \Tr \left[ S_j^{\rho}(p) S_i^{\rho}(p+q) \right] \times \\\notag
    & \times [1+f_B(q_0) - f_F(q_0-k_0)]
    [f_F(p_0) - f_F(p_0+q_0)]
    \,,
\end{align}
where $S^\rho$ are the spectral fermion propagators.
At tree level, for a particle of mass $m_x$, these are given by $S^\rho_x(k) = \pi P_x (\slashed{k}+m_x) \delta (k^2 - m^2_x)\,\mathrm{sign}\,(k_0)$, with a chiral projection operator (in the case of chiral fermions) $P_x$.
For HNL production, we are primarily interested in the relativistic limit, where the temperature is much greater than all the particle masses, so we set $m_x = 0$ for the rest of the computation, and assume that all particles are chiral fermions.

After performing the integrals over the momentum $p$ and the angular $q$-integrals, the production rate simplifies to
\begin{align}
\label{eq:rate_scalar}
k_0 \Gamma = G_F^2 |c_{S,ijk}|^2 A_{S,ijk} \frac{1}{(4 \pi)^3 |k|}
    & \left[
    \int_{k_0}^\infty d q_+ \int_0^{k_0} d q_-
    [1 + f_B(q_0) - f_F(q_0-k_0)] \mathcal{I}_s(q) +
    \right.
    \\\notag
    &\left.
    + \int_{0}^{k_0} d q_+ \int_{-\infty}^0 d q_-
    [1 + f_B(q_0) - f_F(q_0-k_0)] \mathcal{I}_t(q)
    \right]\ ,
\end{align}
where the integrals $\mathcal{I}$ are given by
\begin{align}
    \mathcal{I}_s(q) = 2 T (\ln_\mathrm{f^+} - \ln_\mathrm{f^-}) + q\,,\qquad \mathcal{I}_t(q) = 2 T (\ln_\mathrm{f^-} - \ln_\mathrm{f^+})\,,
\end{align}
with $q_\pm = (q_0 \pm q)/2$, and the notation:
\begin{align}
    \ln_\mathrm{f^-} = \ln( 1 + e^{-|q_-|/T})\,,\qquad  \ln_\mathrm{f^+} = \ln( 1 + e^{-q_+/T})\,.
\end{align}
The remaining momentum integrals are equivalent to those for the vector-mediated channel in Eq.~\eqref{eq:GammaProd}, except for a relative factor $4$ suppression.
Therefore, for scalar-mediated channels, we use:
\begin{align}
    A_{ijk} = \frac{A_{S,ijk}}{4}\,,
\end{align}
together with the production rate from Eq.~\eqref{eq:GammaProd}.

\subsection{Production rate in the leptoquark scenario}
Because HNL production is already active in the symmetric phase of the SM, it is more practical to work directly with the $\nu$SMEFT coupling $C^{(6)}_{LdQ\nu}$. 
This coupling includes only a scalar operator, unlike the full $\nu$LEFT basis, where it is decomposed into multiple scalar and tensor components.
To make direct use of Eq.~\eqref{eq:rate_scalar}, we introduce
\begin{align}
    c_\mathrm{S}^\prime \equiv C^{(6)}_{LdQ\nu} v^2 = 2 c^\mathrm{CC}_\mathrm{SRR} 
    \simeq -2 c^\mathrm{NC}_\mathrm{SRR}\ .
\end{align}
The production rate is then given by
\begin{align}
    \Gamma_N^{\rm prod} \approx 0.3 |c_\mathrm{S}^\prime|^2 G_F^2 T^5 \cdot \frac{2 \cdot 2 \cdot N_c}{4} = 0.3 A_\mathrm{SRR} |c^\mathrm{CC}_\mathrm{SRR}|^2 G_F^2 T^5\ ,
\end{align}
where one factor of $2$ comes from considering Majorana HNLs and another from taking into account both NC and CC interactions, leading to an overall $A_\mathrm{SRR} = 4 N_c$.

\section{Simplified BBN estimate for very long-lived HNLs}
\label{app:MatterDomination}
To verify that HNLs that are stable during BBN still modify the elemental abundances, we perform a simple estimate for the Helium abundance (see e.g.~\cite{Rubakov:2017xzr}).
The Helium abundance is determined by the neutron-proton ratio at the temperature of thermonuclear reactions, i.e., when $T_{\rm NS} \approx 80$ keV.
The neutron-proton ratio sets the fraction of $^4$He:
\begin{align}
    X_{^4\text{He}} = \frac{2}{n_p(T_{\rm NS})/n_n(T_{\rm NS}) + 1}\ .
\end{align}
The neutron-proton abundance is approximately given by:
\begin{align}
    \frac{n_n(T_{\rm NS})}{n_p(T_{\rm NS})} =
    \frac{n_n(T_N) e^{-T_{\rm NS}/\tau_n}}{
    n_p(T_N) + n_n(T_N)(1-e^{-T_{\rm NS}/\tau_n})
    }\ ,
\end{align}
and is therefore extremely sensitive to the exact value of $T_{\rm NS}$.
In the case of radiation domination, this can be obtained from the relation
\begin{align}
    T_{\rm NS} = \frac{1}{2 H} \approx \frac{M_\mathrm{Pl}^*}{2 T_{\rm NS}^2}\,.
\end{align}
However, if a substantial abundance of HNLs is present, it modifies $T_{\rm NS}$.
A rough estimate can be obtained by demanding that $\rho_N < \rho_{SM}$ translates into a bound on the HNL mass $M_4$ through
\begin{align}
    \frac{\rho_N}{\rho_{SM}} = \frac43 \frac{M_4}{T} Y^{f}_N < 1 \rightarrow M_4 < \frac34 \frac{T}{Y_N^f}\,,
\end{align}
using $\rho_N = s_{SM} Y_N^f M_4$ and the SM equation of state.
This corresponds to a bound of about $M_4 < 30$ MeV if we take $T=T_{\rm NS} \approx 80$ keV.

\bibliographystyle{JHEP}
\bibliography{references}

\end{document}